\documentclass[useAMS,usenatbib]{style/mnras}

\usepackage{epsfig,amsmath,natbib,wasysym}
\usepackage{color}

\usepackage{xcolor}
\usepackage{amssymb}
\usepackage{multirow}
\usepackage{framed}
\usepackage{subcaption} 

\usepackage{placeins} % enables \FloatBarrier

\usepackage{etoolbox}
\makeatletter
\newcount\c@additionalboxlevel
\newcount\c@maxboxlevel
\patchcmd\@combinedblfloats{\box\@outputbox}{%
  \stepcounter{additionalboxlevel}%
  \box\@outputbox
}{}{\errmessage{\noexpand\@combinedblfloats could not be patched}}

\AtBeginShipout{%
  \ifnum\value{additionalboxlevel}>\value{maxboxlevel}%
    \typeout{Warning: maxboxlevel might be too small, increase to %
      \the\value{additionalboxlevel}%
    }%
  \fi 
  \@whilenum\value{additionalboxlevel}<\value{maxboxlevel}\do{%
    \typeout{* Additional boxing of page `\thepage'}%
    \setbox\AtBeginShipoutBox=\hbox{\copy\AtBeginShipoutBox}%
    \stepcounter{additionalboxlevel}%
  }%
  \setcounter{additionalboxlevel}{0}%
}
\makeatother

\def\refjnl#1{{\rm#1}}
\def\apjl{\refjnl{ApJ}}                % Astrophysical Journal, Letters
\def\apjs{\refjnl{ApJS}}               % Astrophysical Journal, Supplement
\def\ao{\refjnl{Appl.~Opt.}}           % Applied Optics
\def\aap{\refjnl{A\&A}}                % Astronomy and Astrophysics
\def\refjnl#1{{\rm#1}}
\def\apjl{\refjnl{ApJ}}                % Astrophysical Journal, Letters
\def\apjs{\refjnl{ApJS}}               % Astrophysical Journal, Supplement
\def\ao{\refjnl{Appl.~Opt.}}           % Applied Optics
\def\aap{\refjnl{A\&A}}                % Astronomy and Astrophysics
\def\be{\begin{equation}} 
\def\ee{\end{equation}} 
\def\ba{\begin{eqnarray}} 
\def\ea{\end{eqnarray}}

\def\kms{\,{\rm {km\, s^{-1}}}} 
\def\cc{\,{\rm {cm^{-3}}}} 
\def\msun{{\Msun}}

\def\HH{${\rm {H_2}}\,\,$}

\def\HI{\hbox{H~$\scriptstyle\rm I\ $}} 
\def\HII{\hbox{H~$\scriptstyle\rm II\ $}}

\def\gsim{\lower.5ex\hbox{\gtsima}} 
\def\lsim{\lower.5ex\hbox{\ltsima}} \def\gtsima{$\; \buildrel > \over 
\sim \;$} \def\ltsima{$\; \buildrel < \over \sim \;$} \def\prosima{$\; 
\buildrel \propto \over \sim \;$} \def\gsim{\lower.5ex\hbox{\gtsima}} 
\def\lsim{\lower.5ex\hbox{\ltsima}} 
\def\simgt{\lower.5ex\hbox{\gtsima}} 
\def\simlt{\lower.5ex\hbox{\ltsima}} 
\def\simpr{\lower.5ex\hbox{\prosima}} \def\la{\lsim} \def\ga{\gsim} 
  
 \def\gtsima{$\; \buildrel > \over \sim \;$} 
\def\ltsima{$\; \buildrel < \over \sim \;$} 
\def\gsim{\lower.5ex\hbox{\gtsima}} 
\def\lsim{\lower.5ex\hbox{\ltsima}} 
\def\simgt{\lower.5ex\hbox{\gtsima}} 
\def\simlt{\lower.5ex\hbox{\ltsima}} 
\def\simpr{\lower.5ex\hbox{\prosima}} 
\def\la{\lsim} 
\def\ga{\gsim}

\def\msun{\,{\rm \Msun}}

\def\E3{{\cal E}_{\rm g}^{III}}

\def\Msun{M_\odot}

\def\r12{r_{1/2}} 
\def\x12{x_{1/2}} 
\def\v12{v_{1/2}}

\newcommand{\quotes}[1]{``#1''}

\def\ramses{\textsc{ramses}}
\def\ramsesrt{\textsc{ramses-rt}}
\def\krome{\textsc{krome}}
\def\pymses{\textsc{pymses}}

\def\lsun{{\rm L}_{\odot}}
\def\msun{{\rm M}_{\odot}}

\def\h2{{\mathrm{H}_2}}

\def\hi{\mathrm{HI}}

\def\CII{\hbox{[C~$\scriptstyle\rm II $]}}

\def\pc{\mathrm{pc}}
\def\cms{\mathrm{cm}\,\mathrm{s}^{-1}}
\def\kms{\mathrm{km}\,\mathrm{s}^{-1}}

\definecolor{apcolor}{HTML}{b3003b}
\definecolor{afcolor}{HTML}{800080}
\definecolor{ddcolor}{HTML}{077a2f}

\defcitealias{Decataldo2017}{D17}
\def\D17cit{\citetalias{Decataldo2017}}

\makeatletter
\newcommand\footnoteref[1]{\protected@xdef\@thefnmark{\ref{#1}}\@footnotemark}
\makeatother

\title[Photoevaporation of Jeans-unstable molecular clumps]{Photoevaporation of Jeans-unstable molecular clumps}
\author[Decataldo et al.]{
D. Decataldo$^{1}$\thanks{\href{mailto:davide.decataldo@sns.it}{davide.decataldo@sns.it}},
A. Pallottini$^{2,1}$, 
A. Ferrara$^{1,3}$, 
L. Vallini$^{4,5}$,
S. Gallerani$^{1}$
\\
$^{1}$ Scuola Normale Superiore, Piazza dei Cavalieri 7, I-56126 Pisa, Italy\\
$^{2}$ Centro Fermi, Museo Storico della Fisica e Centro Studi e Ricerche ``Enrico Fermi'', Piazza del Viminale 1, Roma, 00184, Italy\\
$^{3}$ Kavli IPMU, The University of Tokyo, 5-1-5 Kashiwanoha, Kashiwa 277-8583, Japan\\
$^{4}$ Leiden Observatory, Leiden University, PO Box 9500, 2300 RA Leiden, The Netherlands\\
$^{5}$ Nordita, KTH Royal Institute of Technology and Stockholm University, Roslagstullsbacken 23, SE-10691 Stockholm, Sweden
}

\begin{document}
 
\date{\today} 
 
\pagerange{\pageref{firstpage}--\pageref{lastpage}} \pubyear{2019}
 
\maketitle 

\setlength{\parskip}{0pt}
    
\label{firstpage} 
\begin{abstract}
We study the photoevaporation of Jeans-unstable molecular clumps by isotropic FUV ($6\,{\rm eV} < {\rm h}\nu < 13.6\,{\rm eV}$) radiation, through 3D radiative transfer hydrodynamical simulations implementing a non-equilibrium chemical network that includes the formation and dissociation of H$_2$. We run a set of simulations considering different clump masses ($M=10-200$ M$_\odot$) and impinging fluxes ($G_0=2\times 10^3-8\times 10^4$ in Habing units). In the initial phase, the radiation sweeps the clump as an R-type dissociation front, reducing the H$_2$ mass by a factor $40-90\%$. Then, a weak  ($\mathcal{M}\simeq 2$) shock develops and travels towards the centre of the clump, which collapses while loosing mass from its surface. All considered clumps remain gravitationally unstable even if radiation rips off most of the clump mass, showing that external FUV radiation is not able to stop clump collapse. However, the  FUV intensity regulates the final H$_2$ mass available for star formation: for example, for $G_0 < 10^4$ more than 10\% of the initial clump mass survives. Finally, for massive clumps ($\apprge 100$ M$_\odot$) the H$_2$ mass {\it increases} by $25-50\%$ during the collapse, mostly because of the rapid density growth that implies a more efficient H$_2$ self-shielding.\end{abstract}


% ************************************************************************************************ INTRODUCTION

\section{Introduction}
\label{Intro}

Stars are known to form in clusters inside giant molecular clouds (GMCs), as a consequence of the gravitational collapse of overdense clumps and filaments \citep{Bergin1996, Wong2008, Takahashi2013, Schneider2015, Sawada2018}. The brightest (e.g. OB) stars have a strong impact on the surrounding interstellar medium (ISM), since their hard radiation field ionizes and heats the gas around them, increasing its thermal pressure. As a result, the structure of the GMC can be severely altered due to the feedback of newly formed stars residing inside the cloud, with the subsequent dispersal of low-density regions. Collapse can then occur only in dense regions able to self-shield from  impinging radiation \citep{Dale2005, Dale2012a, Dale2012, Walch2012}.

The ISM within the Str\"omgren sphere around a star-forming region is completely ionized by the extreme ultra-violet (EUV) radiation, with energy above the ionization potential of hydrogen ($h\nu > 13.6$ eV). The typical average densities of \HII regions are $\langle n \rangle \simeq 100 \,\cc$: this results in ionization fractions $x_\textsc{hii} < 10^{-4}$ and a final gas temperature $T>10^{4-5}$ K. Far-ultraviolet (FUV) radiation (photon energy $6\,\mathrm{eV} < h\nu < 13.6 \,\mathrm{eV}$) penetrates beyond the \HII region, thus affecting the physical and chemical properties of the ISM up to several parsecs. This region is usually referred to as the Photo-Dissociation Region \citep[PDR;][]{Tielens1985, Kaufman1999, LePetit2006, Bron2018}. Typical fluxes in the FUV band due to OB associations may have values as high as $G_0=10^{4-5}$ \citep{Marconi1998}, in units of the Habing flux\footnote{The Habing flux ($1.6 \times 10^{-3}\, {\rm erg}\, {\rm s}^{-1} {\rm cm}^{-2}$) is the average interstellar radiation field of our Galaxy in the range [6 eV, 13.6 eV] \citep{Habing1968}}. A PDR is characterized by a layer with neutral atomic hydrogen, photo-dissociated by Lyman-Werner photons ($11.2 \,\mathrm{eV} < h\nu < 13.6\, \mathrm{eV}$), and a deeper layer where gas self-shielding allows hydrogen to survive in molecular form. There are many observational pieces of evidence that the structure of PDRs are not homogeneous, with gas densities spanning many orders of magnitude from $10^2\,\cc$ to $10^6\,\cc$. In particular, small isolated cores of few solar masses and sizes $\apprle$ 0.1 pc are commonly observed \citep{Reipurth1983, Hester1996, Huggins2002, Makela2013}. Radiative feedback by FUV radiation could explain their formation via shock-induced compression \citep{Lefloch1994}.

The effect of FUV radiation on clumps is twofold: (1) FUV radiation dissociates the molecular gas, which then escapes from the clump surface at high velocity (photoevaporation); (2) radiation drives a shock which induces the clump collapse \citep[radiation-driven implosion, RDI][]{Sandford1982}. The first effect reduces the clump molecular mass, hence decreasing the mass budget for star formation within the clump. Instead, the latter effect may promote star formation by triggering the clump collapse. Hence, the net effect of radiative feedback on dense clumps is not trivial and deserves a careful analysis.

The flow of gas from clumps immersed in a radiation field has been studied by early works both theoretically \citep{Dyson1968, Mendis1968a, Kahn1969, Dyson1973} and numerically \citep{Tenorio-Tagle1977, Bedijn1984}. \citet{Bertoldi1989} and \citet{Bertoldi1990} developed semi-analytical models to describe the photoevaporation of atomic and molecular clouds induced by ionizing radiation. In their models they also include the effects of magnetic fields and self-gravity. They find that clumps settle in a stationary cometary phase after the RDI, with clump self-gravity being negligible when the magnetic pressure dominates with respect to the thermal pressure (i.e. $B>6\,\mu \mathrm{G}$), or when the clump mass is much smaller than a characteristic  mass $m_\mathrm{ch}\simeq 50 \,\msun$. They focused on gravitationally stable clumps, thus their results are not directly relevant for star formation.

Later, the problem was tackled by \citet{Lefloch1994}, who performed numerical simulations which however only included the effect of thermal pressure on clump dynamics.
Gravity was then added for the first time by \citet{Kessel-Deynet2003}. For an initially gravitationally stable clump of 40 $\msun$, they find that the collapse can be triggered by the RDI; nevertheless, they notice that the collapse does not take place if a sufficient amount of turbulence is injected ($v_{\rm rms} \simeq 0.1 \,{\rm km/s}$).
\citet{Bisbas2011} also ran simulations of photoevaporating clumps, with the specific goal of probing triggered star formation. They find that star formation occurs only when the intensity of the impinging flux is within a specific range ($10^9 \,{\rm cm}^{-2} {\rm s}^{-1} < \Phi_\textsc{euv} < 3\times10^{11} \,{\rm cm}^{-2} {\rm s}^{-1}$ for a $5\,\msun$ initially stable clump
\footnote{Assuming for example the spectrum of a $10^4\,\lsun$, this EUV flux corresponds roughly to a flux $G_0=10-3\times 10^4$ in the FUV band.}
). All these works include the effect of ionizing radiation only, while FUV radiation feedback is instead relevant for clumps located outside the \HII region of a star (cluster).

In a previous work \citep[hereafter \D17cit]{Decataldo2017}, we have constructed a 1D numerical procedure to study the evolution of a molecular clump, under the effects of both FUV and EUV radiation. We have followed the time evolution of the structure of the iPDR (ionization-photodissociation region) and we have computed the photoevaporation time for a range of initial clump masses and intensity of impinging fluxes. However, since \D17cit did not account for gravity, $\h2$ dissociation is unphysically accelerated during the expansion phase following RDI. Those results have been compared with the analytical prescriptions by \citet{Gorti2002}, finding photoevaporation times in agreement within a factor 2, although different simplifying assumptions where made in modelling the clump dynamics.

The same setup by \D17cit has been used by \citet{Nakatani2018} to run 3D simulations with on-the-fly radiative transfer (RT) and a chemical network including $\mathrm{H}^{+}$, $\h2$, $\mathrm{H}^{+}$, $\mathrm{O}$, $\mathrm{CO}$ and $\mathrm{e}^{-}$. Without the inclusion of gravity in their simulations, they find that the clump is confined in a stable cometary phase after the RDI, which lasts until all the gas is dissociated and flows away from the clump surface. Nevertheless, they point out that self-gravity may affect the clump evolution when photoevaporation is driven by a FUV-only flux, while the EUV radiation produces very strong photoevaporative flows which cannot be suppressed by gravity.

In this paper, we attempt to draw a realistic picture of clump photoevaporation by running 3D hydrodynamical simulations with gravity, a non-equilibrium chemical network including formation and photo-dissociation of $\h2$, and an accurate RT scheme for the propagation of FUV photons. We focus on the effect of radiation on Jeans-unstable clumps, in order to understand whether their collapse is favoured or suppressed by the presence of nearby stars emitting in the FUV range. 

The paper is organized as follows. In Sec. \ref{Met}, we describe the numerical scheme used for the simulations, and, in Sec. \ref{computational_box}, the initial conditions for the gas and radiation. We analyse the evolution of the clump for different radiative fluxes and masses is Sec. \ref{Res_M50} and Sec. \ref{Res_Mdiff}, respectively. A cohesive picture of the photoevaporation process is given in Sec. \ref{Res_final}. Our conclusions are finally summarized in Sec. \ref{Con}.

% ************************************************************************************************ METHODS

\section{Numerical scheme}  \label{Met}        % ------------------- SEC : NUMERICAL SUITE

\begin{figure}
\centering
\includegraphics[width=0.45\textwidth]{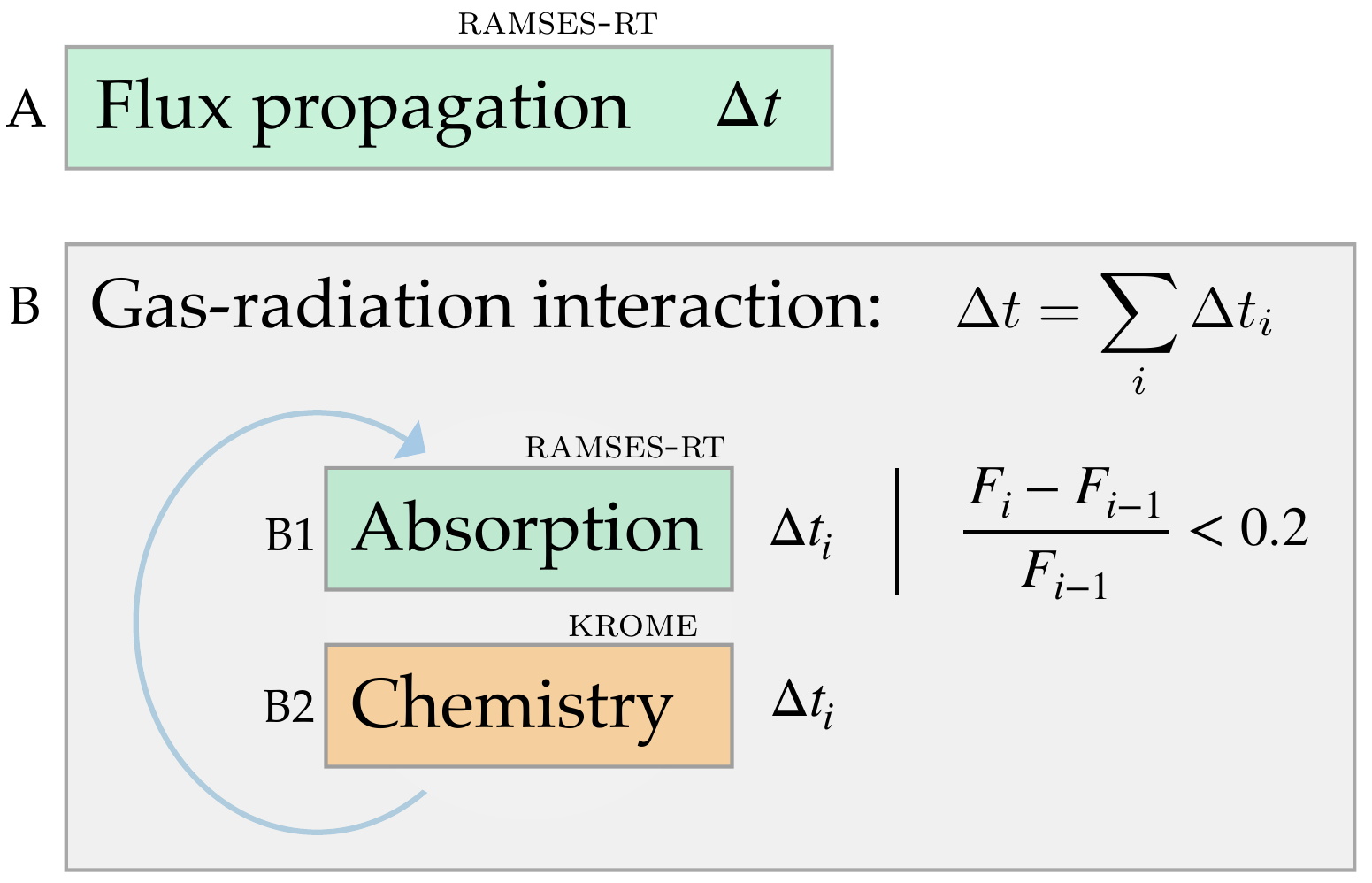}
\caption{Diagram of the $\ramsesrt$ and $\krome$ coupling.
In each time step $\Delta t$, first the flux $F$ is propagated without attenuation (A). Then, the gas-radiation interaction step is carried out (B), by sub-cycling in radiation absorption (B1) and chemical evolution (B2) steps. Each sub-step is evolved for a time $\Delta t_i$, that is chosen to assure a fractional variation $<$20\% for the impinging flux.
See text for details.
\label{ramses_krome}}
\end{figure}

\begin{table}
\centering
\begin{tabular}{ll}
\multicolumn{2}{c}{Photochemical reactions}                                                                                            \\ \hline
${\rm H}+\gamma \rightarrow {\rm H}^+ + e$         & \quad${\rm H}_2^+ +\gamma \rightarrow {\rm H}^+ + {\rm H}$         \\
${\rm He}+\gamma \rightarrow {\rm He}^+ + e$       & \quad${\rm H}_2^+ +\gamma \rightarrow {\rm H}^+ + {\rm H}^+ + e$   \\
${\rm He}^+ +\gamma \rightarrow {\rm He}^{++} + e$ & \quad${\rm H}_2^+ +\gamma \rightarrow {\rm H} + {\rm H}\,\,$ (direct)  \\
${\rm H}^- +\gamma \rightarrow {\rm H} + e$        & \quad${\rm H}_2^+ +\gamma \rightarrow {\rm H} + {\rm H}\,\,$ (Solomon) \\
${\rm H}_2 +\gamma \rightarrow {\rm H}_2^+ + e$    &                                                                                  
\end{tabular}
\caption{List of photochemical reactions included in our chemical network.
\label{photochemical_reactions}
}
\end{table}

Our simulations are carried out with $\ramsesrt$\footnote{\url{https://bitbucket.org/rteyssie/ramses}}, an adaptive mesh refinement (AMR) code featuring on-the-fly RT \citep{Teyssier2002, Rosdahl2013}. RT is performed with a momentum-based approach, using a first-order Godunov solver and the M1 closure relation for the Eddington tensor. The basic version of $\ramsesrt$ thermochemistry module accounts only for the photoionization of hydrogen and (first and second) photoionization of helium. We have used the chemistry package $\krome$ \footnote{\url{https://bitbucket.org/tgrassi/krome}} \citep{Grassi2014} to implement a complete network of H and He reactions, including neutral and ionized states of He, H and $\h2$.

We track the time evolution of the following 9 species: H, $\mathrm{H}^{+}$, $\mathrm{H}^{-}$, $\mathrm{H}_2$, $\mathrm{H}_2^{+}$, He, $\mathrm{He}^{+}$, $\mathrm{He}^{++}$ and free electrons. Our  chemical network includes 46 reactions in total\footnote{The included reactions and the respective rates are taken from \citet{Bovino2016}: reactions 1 to 31, 53, 54 and from 58 to 61 in Tab. B.1 and B.2, photoreactions P1 to P9 in Tab. 2.}, and features neutral-neutral reactions, charge-exchange reactions, collisional dissociation and ionization, radiative association reactions and cosmic ray-induced reactions (we consider a cosmic ray ionization rate $\zeta_\mathrm{H}=3\times 10^{-17} \,{\rm s}^{-1}$, the reference value in the Milky Way \citep{Webber1998}. We follow $\h2$ formation on dust, adopting the Jura rate at solar metallicity $R_f(\h2)=3.5 \times 10^{-17} n_\mathrm{H} n_\mathrm{tot} $ \citep{Jura1975}. There are 9 reactions involving photons, listed in Tab. \ref{photochemical_reactions}: photoionization of H, He, $\mathrm{He}^{+}$, $\mathrm{H}^{-}$ and $\mathrm{H}_2$ to $\mathrm{H}_2^{+}$, direct photodissociation of $\mathrm{H}_2^{+}$ and  $\mathrm{H}_2$ and the two-step Solomon process \citep{Draine1996a}. The Solomon process rate is usually taken to be proportional to the total flux at 12.87 eV \citep{Glover2007, Bovino2016}, but this is correct only if the flux is approximately constant in the Lyman-Werner band (11.2 - 13.6 eV), as pointed out by \citet{Richings2014a}. These authors find that the most general way to parametrize the dissociation rate is 
\be
\Gamma_\h2 = 7.5 \times 10^{-11} \dfrac{n_{\gamma}(12.24-13.51\,\mathrm{eV})}{2.256\times10^4 \rm cm^{-3}} \, \mathrm{s}^{-1} \,,  
\ee
where $n_{\gamma}(\Delta E_\mathrm{bin})$ is the photon density in the energy interval $\Delta   E_\mathrm{bin}$. We work in the on-the-spot approximation, hence photons emitted by recombination processes are neglected.

Given the chemical network and the included reactions, the code can in principle be used with an arbitrary number of photon energy bins. In the particular context of photoevaporating clumps, we decided to make only use of two bins with energies in the FUV domain, i.e. [6.0 eV, 11.2 eV] and [11.2 eV, 13.6 eV]. As we consider molecular clumps located outside stellar \HII regions, we expect that EUV radiation does not reach the surface of the clump. On the other hand, we neglect photons with energies $ <6.0$ eV since they do not take part in any chemical reactions of interest in our case.

Fig. \ref{ramses_krome} summarizes the approach we used to couple RT module in $\ramses$ with the non-equilibrium network adopted via $\krome$ \citep[as also done in][]{Pallottini2019}. At each time-steps ($\Delta t$), photons are first propagated from each cell to the nearest ones by $\ramsesrt$ (A). Then, the gas-radiation interaction step (B) is executed, sub-cycling in absorption (B1) and chemical evolution (B2) steps with a time-steps $\Delta t_i<\Delta t$, such that the flux is not reduced by more than $20\%$ at each substep ($\Delta t_i$).

In step B1, we account for (1) photons that take part in chemical reactions, (2) $\mathrm{H}_2$ self-shielding and (3) dust absorption. The optical depth of a cell in the radiation bin $i$ (excluding the Solomon process) is computed by summing over all photo-reactions:
\be
\tau_i =\sum_j n_j \Delta x_\mathrm{cell} \sigma_{ij}   \,\, ,
\ee
where $n_j$ is the number density of the photo-ionized/dissociated species in the reaction $j$, $\Delta x_{\rm cell}$ is the size of the cell, and $\sigma_{ij}$ is the average cross section of the reaction $j$ in the bin $i$. For the Solomon process, the self-shielding factor $S_\mathrm{self}^{\mathrm{H}_2}$ is taken from \citet{Richings2014b}, and it is related to the optical depth by $\tau_\mathrm{self}^{\mathrm{H}_2} = -\log(S_\mathrm{self}^{\mathrm{H}_2}) / \Delta x_\mathrm{cell}$. Absorption from dust is included, with opacities taken from \citet{Weingartner2001}. We have used the Milky Way size distribution for visual extinction-to-reddening ratio $R_V =3.1$, with carbon abundance (per H nucleus) $b_C=60$ ppm in the log-normal populations\footnote{\url{https://www.astro.princeton.edu/~draine/dust/dustmix.html}}.

After every absorption substep, $\krome$ is called in each cell (step B2): photon densities in each bin are passed as an input, together with the current chemical abundances and the gas temperature in the cell, and $\krome$ computes the new abundances after a time-steps accordingly.

In App. \ref{app_tests}, we show two successful tests that we performed to validate our scheme for the coupling between $\ramsesrt$ and $\krome$:
\begin{itemize}
\item[\ref{app_hii}] An ionized region, comparing the results with the analytical solution;
\item[\ref{app_pdr}] The structure of \HH in a PDR, compared with the standard benchmarks of \citet{Rollig2007}.
\end{itemize}

\section{Simulation setup}                % ------------------- SUBSEC : COMPUTATIONAL BOX
\label{computational_box}

\subsection{Gas}\label{sec_gas_ic}   %%%%%% SUBSEC: GAS

\begin{figure}
\centering
\includegraphics[width=0.5\textwidth]{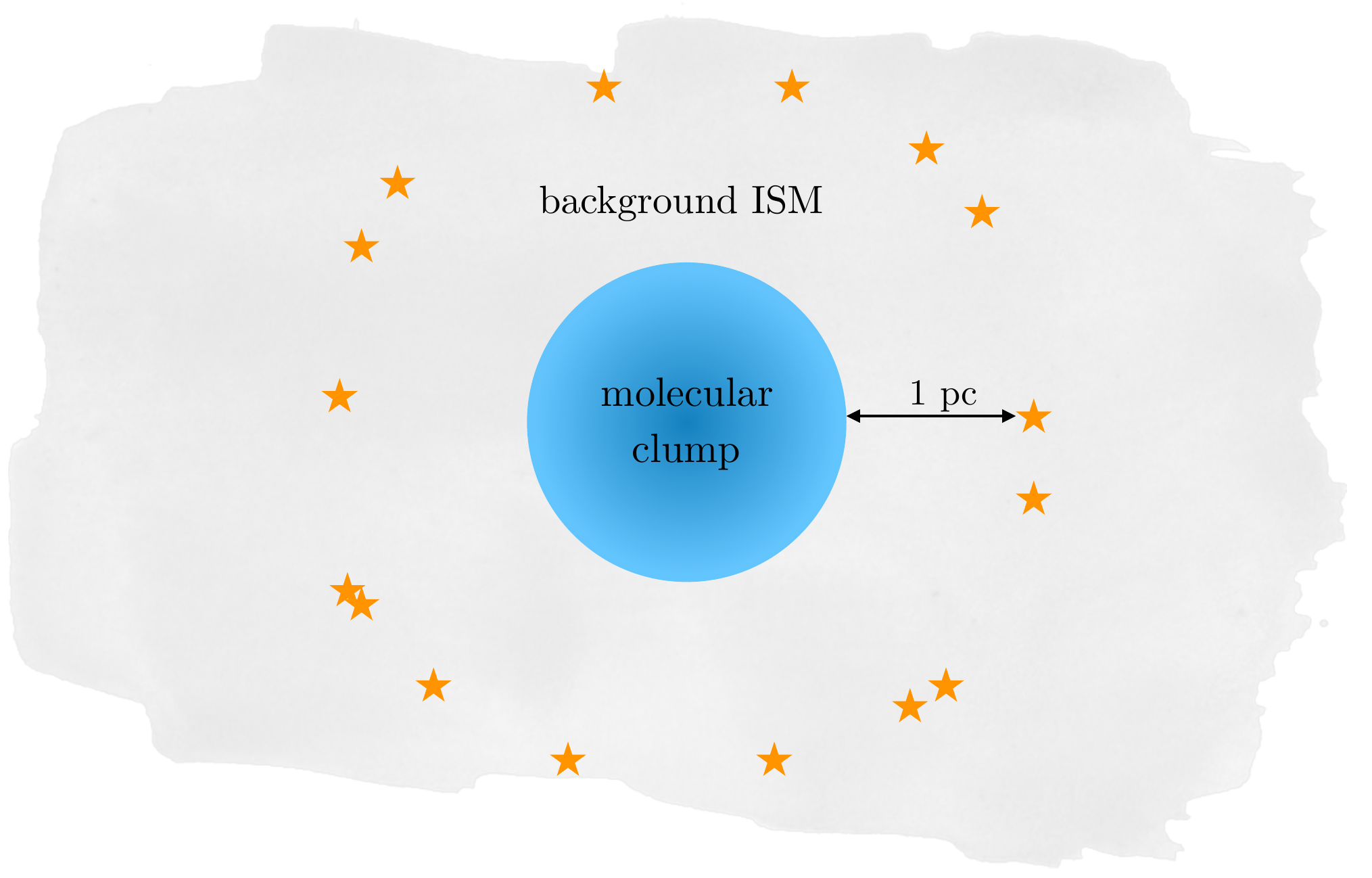}
\caption{Sketch of the simulation set-up. The clump is located at the centre of a box with size 6 pc, filled with a background medium with number density 100 $\cc$. 50 stars are placed at a distance of 1 pc from the surface of the clump, randomly distributed on the surface of a sphere.\label{sketch_ic}}
\end{figure}

\begin{figure}
\includegraphics[width=0.49\textwidth]{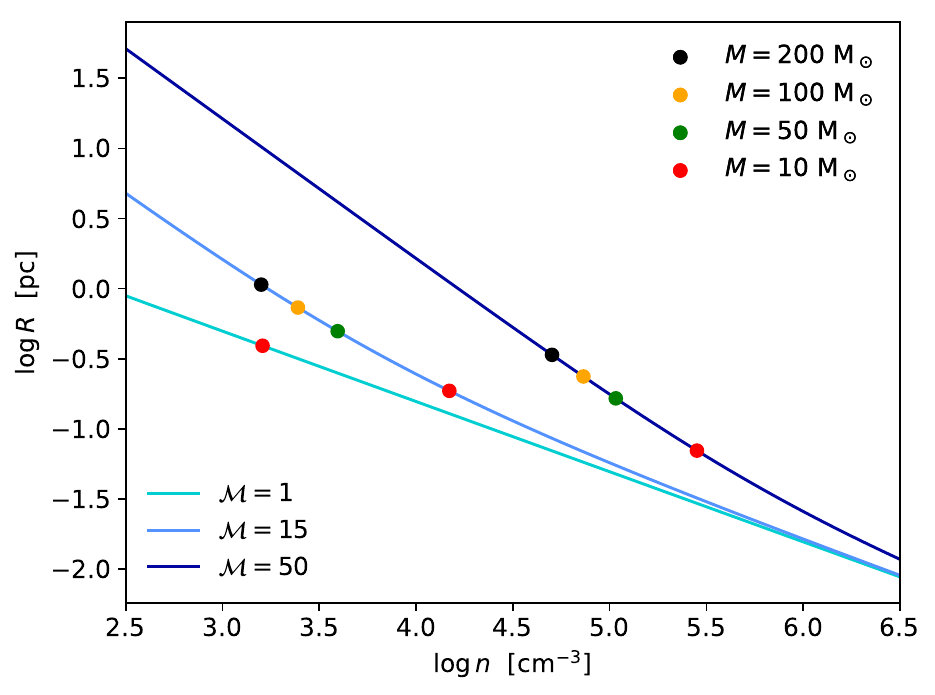}
\caption{Relation between the radius $R$ and the number density $n$ of clumps residing in the parent Giant Molecular Cloud (GMC) with different Mach numbers $\mathcal{M}$. The GMCs have all the same size $L=25$ pc and temperature $T=10$ K. For each GMC, the position in the diagram of clumps with different masses is shown with coloured points.\label{r_n_relation}}
\end{figure}

\begin{figure}
\includegraphics[width=0.49\textwidth]{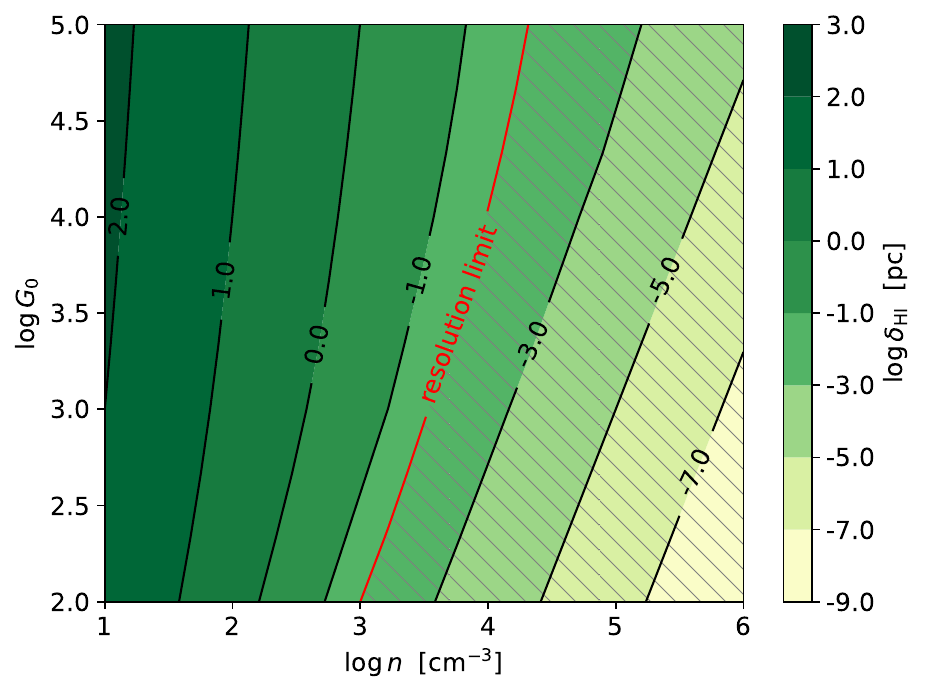}
\caption{Depth of the \HI/$\h2$ transition ($\delta_\textsc{hi}$) in a slab of molecular gas, as a function of the gas number density ($n$) and the FUV flux ($G_0$). The red solid line marks the maximum resolution of our simulations $\Delta x = 0.023$ pc (corresponding to $2^8$ cells). The hatched region highlights the portion of the $G_0$-$n$ diagram where our simulations would not be able to properly resolve $\delta_\textsc{hi}$.\label{layer_thickness}}
\end{figure}

\begin{table*}
\centering
\begin{tabular}{ccccccccc}
\hline \hline
                                       			 & $M\,[\msun]$  & $R\,[\pc]$  & $\langle n \rangle \, [\cc]$  & $n_c \, [\cc]$     & $t_{\rm ff} \, [{\rm Myr}]$  & $G_0$              \\ \hline
\multicolumn{1}{l|}{\texttt{clump\_M50\_noRad}}  & 50            & 0.5		   & $3.9\times 10^3$			   & $6.6\times 10^3$	& 0.81                         & 0                  \\ \hline   
\multicolumn{1}{l|}{\texttt{clump\_M50\_G2e3}}   & 50            & 0.5		   & $3.9\times 10^3$			   & $6.6\times 10^3$	& 0.81                         & $2 \times 10^3$    \\
\multicolumn{1}{l|}{\texttt{clump\_M50\_G3e4}}   & 50            & 0.5		   & $3.9\times 10^3$			   & $6.6\times 10^3$	& 0.81                         & $3 \times 10^4$    \\
\multicolumn{1}{l|}{\texttt{clump\_M50\_G8e4}}   & 50            & 0.5		   & $3.9\times 10^3$			   & $6.6\times 10^3$	& 0.81                         & $8 \times 10^4$    \\ \hline
\multicolumn{1}{l|}{\texttt{clump\_M10\_G3e4}}   & 10            & 0.2		   & $1.5\times 10^4$			   & $2.1\times 10^4$	& 0.41                         & $3 \times 10^4$    \\
\multicolumn{1}{l|}{\texttt{clump\_M100\_G3e4}}  & 100           & 0.7		   & $2.5\times 10^3$			   & $4.4\times 10^3$	& 1.01                         & $3 \times 10^4$    \\ 
\multicolumn{1}{l|}{\texttt{clump\_M200\_G3e4}}  & 200           & 1.0		   & $1.6\times 10^3$			   & $3.3\times 10^3$	& 1.26                         & $3 \times 10^4$    \\  \hline 
\end{tabular}

\caption{Summary of the 3D simulation run in this work. Given a mass $M$, the corresponding radius $R$ and density (average number density $\langle n \rangle$ and central number density $n_c$) are determined, as detailed in Sec. \ref{sec_gas_ic}. Simulations of clumps with the same mass differ for the intensity of the external source of FUV radiation $G_0$, that is calculated at the clump surface (see Sec. \ref{rad_sources}). The free-fall time is also reported for reference.
\label{simulation_list}
}
\end{table*}

The computational box is filled with molecular gas\footnote{The gas has helium relative mass abundance $X_{\rm He} = 25\%$.} of density $n = 100\,\cc$ and metallicity $Z={\rm Z}_{\odot}$. A dense clump ($n=10^3-10^4\,\cc$) is then located at the centre of the domain, with the same initial composition of the surrounding gas (Fig. \ref{sketch_ic}). 

Clumps in GMCs are self-gravitating overdensities.  Observations of GMCs of different sizes and masses \citep{Hobson1992, Howe2000, Lis2003, Minamidani2011, Parsons2012, Barnes2018, Liu2018} show that clumps have a wide range of physical properties: radii range from $0.1-10$ pc, densities can be $10-10^4$ times the average density of the GMC; typical masses range from few solar masses to few hundreds $\msun$. 

The density distribution of clumps in GMCs has been studied with numerical simulations of supersonic magnetohydrodynamic turbulence \citep{Padoan2002, Krumholz2005a, Padoan2011, Federrath2013}, yielding a log-normal PDF (Probability Distribution Function):
\begin{subequations}
\be
g(s) = \dfrac{1}{(2 \pi \sigma^2)^{1/2}} \, \exp \left[ -\dfrac{1}{2} \, \left( \dfrac{s-s_0}{\sigma} \right) \right]    \,,
\ee 
where $s = \ln(n / n_0)$, with $n_0$ being the mean GMC density, $s_0 = -\sigma^2 / 2$, and $\sigma$ a parameter quantifying the pressure support. Turbulent and magnetic contribution to gas pressure can be parametrized by the Mach number $\mathcal{M}$ and the thermal-to-magnetic pressure ratio $\beta$, respectively; then $\sigma$ is given by
\be
\sigma^2 = \ln \left( 1 + b^2 \mathcal{M}^2 \dfrac{\beta}{\beta + 1} \right)     \,,
\ee
\end{subequations}
where $b\simeq 0.3-1$ is a factor taking into account the kinetic energy injection mechanism which is driving the turbulence \citep{Molina2012}. Including self-gravity, the high-density end of the PDF is modified with a power-law tail $g(n) \sim n^{-\kappa}$, with $\kappa\sim 1.5-–2.5$ \citep{Krumholz2005a, Padoan2011, Federrath2013, Schneider2015}.

If a value $n$ of the density is drawn from the PDF $g(s)$, the corresponding radius of the clump can be estimated with the turbulent Jeans length \citep{Federrath2012}:
\be
R = \dfrac{1}{2}\lambda_\textsc{j} = \dfrac{1}{2} \dfrac{\pi \sigma^2 + \sqrt{36\pi c_s^2 GL^2 m_p \mu n + \pi^2\sigma^4}}{6GL m_p \mu n}
\ee
where $c_s$ is the isothermal sound speed, $L$ is the size of the GMC, $m_p$ the proton mass and $\mu$ the mean molecular weight. The corresponding clump mass is estimated by assuming a spherical shape and uniform density.

In Fig. \ref{r_n_relation} the clump radius is plotted as a function of the number density, for different Mach numbers $\mathcal{M}$, GMC size fixed at $L=25$ pc and $c_s$ computed with a temperature $T=10$ K and molecular gas ($\mu = 2.5$). The coloured points correspond to the position of clumps with different masses $M=10-200 \, \msun$ in the $R$-$n$ diagram. In GMCs with the same Mach number, clumps with larger mass are less dense: indeed, as a rough approximation, we have that $R\propto n^{-1}$ and $M \sim m_p \mu n R^3 \propto n^{-2}$. It is also interesting to check how the properties of clumps with the same mass vary for different parent GMCs. Considering the 10 $\msun$ clump (red point), we notice that decreasing $\mathcal{M}$, its position in the $R$-$n$ diagram shifts towards lower density and larger radius. Thus, in a $\mathcal{M}=1$ cloud, clumps with mass higher than 10 $\msun$ would have a density as low as the average cloud density, implying that clumps more massive than 10 $\msun$ do not exist at all in such a cloud.

Here we consider clumps residing in a GMC with size $L=25$ pc, average temperature $T=10$ K and Mach number $\mathcal{M}=15$. In particular, we explore the range of masses represented in  Fig. \ref{r_n_relation}, i.e. $M=10, 50, 100, 200 \, \msun$. Masses, radii and average densities of these clumps are summarized in Tab. \ref{simulation_list}. Clumps are modelled as spheres located at the centre of the computational box; their initial density profile is constant up to half of the radius, and then falls as a power law:
 \be
  n(r) = \begin{cases}
             n_c                           & \quad     r < R / 2                \\
             n_c(2r/R)^{-1.5}        & \quad     R / 2  \leq r < R   \\
             \end{cases}
.\
\label{clump_profile}
\ee
where for each clump the core density $n_c$ is chosen such that the total mass is the selected one (Tab. \ref{simulation_list}). The profile in Eq. \ref{clump_profile} has been used in simulations of molecular clouds \citep[e.g.][]{Krumholz2011} and it is physically motivated by observations of star-forming clumps \citep{Beuther2002, Mueller2002}. The choice of this profile implies that there is a discontinuity in the gas density (and so the pressure) at the interface with the external ISM. In the simulations, this will produce a slight expansion of the clump just before the stellar radiation hits the clump surface.

A turbulent velocity field is added to the clump in the initial condition. We generate an isotropic random Gaussian velocity field with power spectrum $P(k) \propto k^{-4}$ in Fourier space, normalising the velocity perturbation so that the virial parameter $\alpha = 5 \,v_{\rm rms}^2 \, R_c / G\,M_c$ is equal to 0.1, as measured for some clumps with mass around $10^2\,\msun$ \citep[e.g.][]{Parsons2012}. In three dimensions, the chosen power spectrum gives a velocity dispersion that varies as $\ell^{1/2}$ with $\ell$ the length scale, which is in agreement with Larson scaling relations \citep{Larson1981}.

\subsection{Radiation sources}\label{rad_sources} %%%%%% SUBSEC: RADIATION

We set-up a roughly homogeneous radiation field around the clump by placing 50 identical point sources (i.e. stars), randomly distributed on the surface of a sphere centred on the clump and with radius larger than the clump radius ($R_\mathrm{sources} = R_\mathrm{clump} + 1\,\pc$)\footnote{The choice of 1 pc as a distance of the sources from the clump is arbitrary. In fact the aim is to get a specific $G_0$ at the clump surface, which can be obtained either varying the source luminosity or the source distance. The number of sources is also not relevant, provided that it is large enough to ensure a nearly isotropic flux on the clump surface.}, as depicted in Fig. \ref{sketch_ic}. Each star has a black-body spectrum and a bolometric luminosity in the range $L_\star = 10^4-10^6 \, \lsun$, according to the desired FUV flux at the clump surface (Tab. \ref{simulation_list}).

In our simulations we use the GLF scheme\footnote{To solve numerically the propagation of photons, different function for the intercell photon flux can be used. \ramses~implements the Harten-Lax-van Lee (HLL) function \citep{Harten1983, Gonzalez2007} and the Global Lax Friedrich (GLF), obtained by setting the HLL eigenvalues to the speed of light.\label{RT_schemes}} for the propagation of photons, since it is more suitable for isotropic sources (while the HLL scheme\footnoteref{RT_schemes} introduces asymmetries, see \citealt{Rosdahl2013}). Our configuration of sources is prone to the problem of \quotes{opposite colliding beams}: the photon density is higher than expected in the cells where two or more fluxes come from opposite directions. This problem is due to the M1 closure relation (see \citealt{Gonzalez2007, Aubert2008} and \citealt{Rosdahl2013}, in particular their Fig. 1) and it is detailed in App. \ref{app_oppositeflux}. To circumvent this issue, we compute the average flux on the clump surface at the beginning of the simulation, which could be higher than that expected from an analytical calculation. In Tab. \ref{simulation_list} we list the resulting FUV flux at the clump surface, for the different set-ups.

\subsection{Resolution} %%%%%% SUBSEC: RESOLUTION

The coarse grid has a resolution of $128^3$ cells, which implies a cell size $\Delta x_\mathrm{cell} \simeq 0.047$ pc. We include one AMR level according to a refinement criterion based on the $\h2$ abundance gradients: a cell is refined if the $\h2$ abundance gradient with neighbouring cells is higher than $10\%$. Thus the effective resolutions is increased up to $256^3$ cells with size $\Delta x_\mathrm{cell} \simeq 0.023$ pc. For the control run without radiation, the resolution is increased by 4 additional levels of refinement in a central region of radius 0.05 pc.

The expected time-scale of photoevaporation is of the order of 1 Myr \citep{Gorti2002,Decataldo2017}. Simulations are carried out with a reduced speed of light $c_\mathrm{red}=10^{-3} \, c$, where $c$ is the actual speed of light, in order to prevent exceedingly small time-stepss, which would result in a prohibitively long computational time. Indeed, time-stepss are settled by the light-crossing time of cells in the finest grid, hence in our simulations $\Delta t \simeq  75$ yr with reduced speed of light.
The reduced speed of light affects the results of simulations when $c_\mathrm{red}$ is lower than the speed of ionization/dissociation fronts \citep{Deparis2019, Ocvirk2019}, which is given by $v_\mathrm{front} = \phi / n$, where $\phi$ is the photons flux in cm$^{-2}$s$^{-1}$. In our set of simulations, \texttt{clump\_M50\_G8e4} has the highest $G_0 / \langle n \rangle$ ratio, yielding 
\begin{equation}
v_\mathrm{front} = \dfrac{\phi}{n} \simeq \dfrac{1.6\times 10^{-3} G_0 }{\langle n \rangle \langle h \nu \rangle_\textsc{fuv}} \simeq 10^9 \,\cms > c_\mathrm{red}
\label{DF_speed}
\end{equation}
This points out that the propagation of the dissociation front (DF) is not treated accurately in our simulations. Nevertheless, the DF propagates at $v_\mathrm{front}$ only for a time around few kyr, after which the front stalls and the photoevaporation proceeds for about 0.1-1 Myr. Hence, most of the simulation is not influenced by the reduced speed of light approximation, and the error concerns only the speed of the dissociation front (and not the thermochemical properties of the photo-dissociated gas).

We have checked that the resolution of our simulation is sufficient to describe the effect of radiation on dense gas. In fact, FUV radiation dissociates and heats a shell of gas at the clump surface, hence the simulation is physically accurate only if the thickness of this layer is resolved. The thickness of the dissociated shell corresponds to the depth $\delta_\textsc{hi}$ of the \HI/$\h2$ transition in a PDR. An analytic approximation for the column density of the transition ($N_\mathrm{trans}=\delta_\textsc{hi} n_\textsc{h}$) is given by the expressions \citep{Bialy2016}
\be
N_\mathrm{trans} = 0.7 \, \ln \left[ \left( \dfrac{\alpha G}{2} \right)^{1/0.7} + 1 \right] \left(1.9\times 10^{-21} \, Z \right)^{-1}\,\mathrm{cm}^{-2}
\ee
\be
\alpha G = 0.35 G_0 \left( \dfrac{100 \, \cc}{n_\textsc{h}} \right) \left(\dfrac{9.9}{1+8.9\,Z/{\rm Z}_{\odot}} \right)^{0.37}   \,.
\ee
This expression is obtained by considering the $\h2$ formation-dissociation balance in a slab of gas with constant density, accounting for $\h2$ self-shielding \citep{Sternberg2014} and dust absorption, without the effect of cosmic rays.

In Fig. \ref{layer_thickness}, we have plotted $\delta_\textsc{hi}$ as a function of the FUV flux and the gas density. The SED of the impinging radiation is that of a $10^4 \, \lsun$ star, with a distance scaled to obtain the flux $G_0$ shown in the $x$ axis of the plot. The red line marks the contour corresponding to the maximum resolution of our simulations, i.e. $\Delta x = 0.023$ pc.
Hence, referring to Tab. \ref{simulation_list} for the values of the mean density, $\langle n \rangle$, we can see that the effect of radiation on clumps with mass larger than 50 $\msun$ is well resolved for every value of $G_0$ of interest. Simulations with $M=10, 50\msun$ are close to the resolution limit if $G_0 < 10^4$. Keep in mind, though, that due to the imposed profile Eq. \ref{clump_profile}, the density in the outer regions is smaller than $\langle n \rangle$.

\subsection{Set of simulations}

We run in total seven 3D simulations of photoevaporating dense molecular clumps (Tab. \ref{simulation_list}). The first simulation of a 50 $\msun$ clump does not include radiation and it is used for comparison. Then we run a set of simulations of clumps with the same mass ($M=50\,\msun$) and different intensities of the radiation field ($G_0=2\times 10^3 - 8\times 10^4$). Finally, in the last set of simulations, we consider clumps with masses varying in the range $M=10-200\,\Msun$ and constant impinging radiation ($G_0\simeq 3\times 10^4$).

% ************************************************************************************************ RESULTS

\section{Fiducial  clumps (50 $\msun$)}\label{Res_M50}         % ------------------- SEC : 50 Msun clumps

\begin{figure*}
\centering
\includegraphics[width=\textwidth]{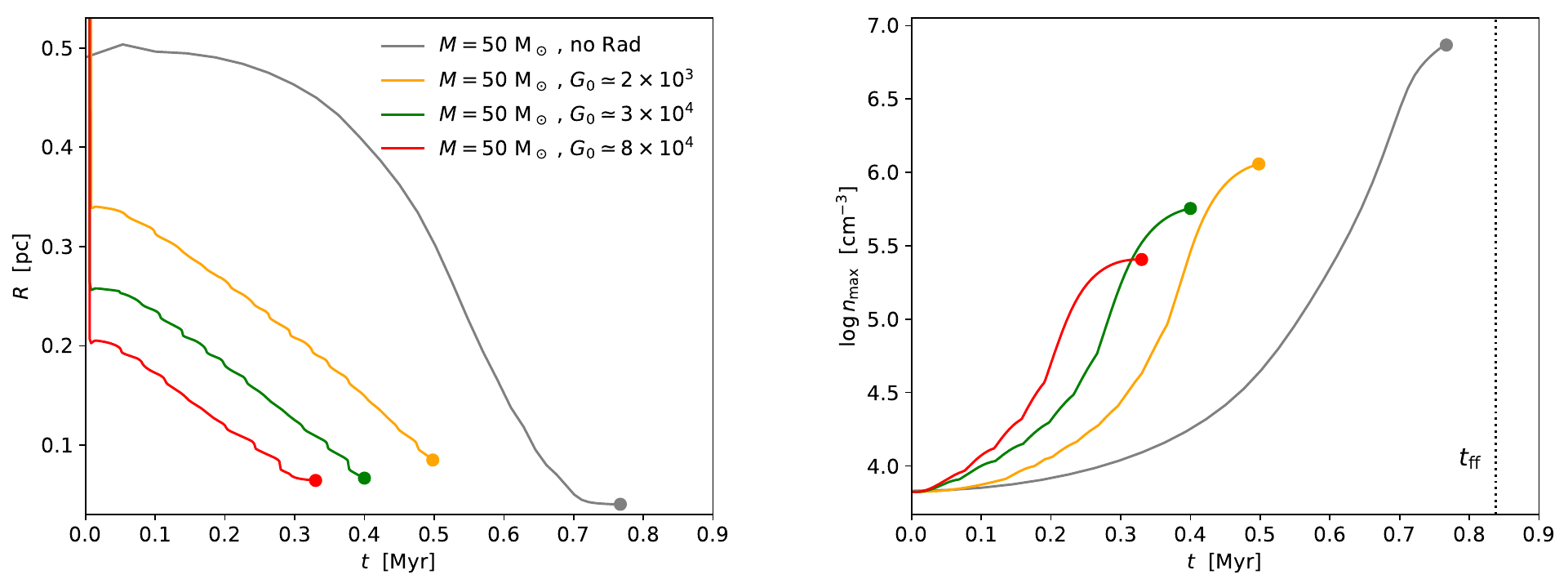}
\caption{Comparison of the evolution of the $50 \, \Msun$ clump in the four simulations without radiation and with $G_0\simeq 2\times 10^3$, $3\times 10^4$, $8\times 10^4$. {\bf Left}: variation with time of the clump radius. The radius is defined as the distance from the centre where 99.7\% of the molecular gas mass is enclosed (apart from \texttt{clump\_M50\_noRad}, where it is defined as the distance from the centre where the density drops to 10\% of the maximum value). {\bf Right}: variation with time of the clump maximum density, with circles marking the time when the clump reaches its minimum radius, and gravitational collapse begins. The black dotted line marks the free fall time, computed with the initial average clump density. \label{clump_r_n_M50}}
\end{figure*}

\begin{figure*}
\centering
\includegraphics[width=\textwidth]{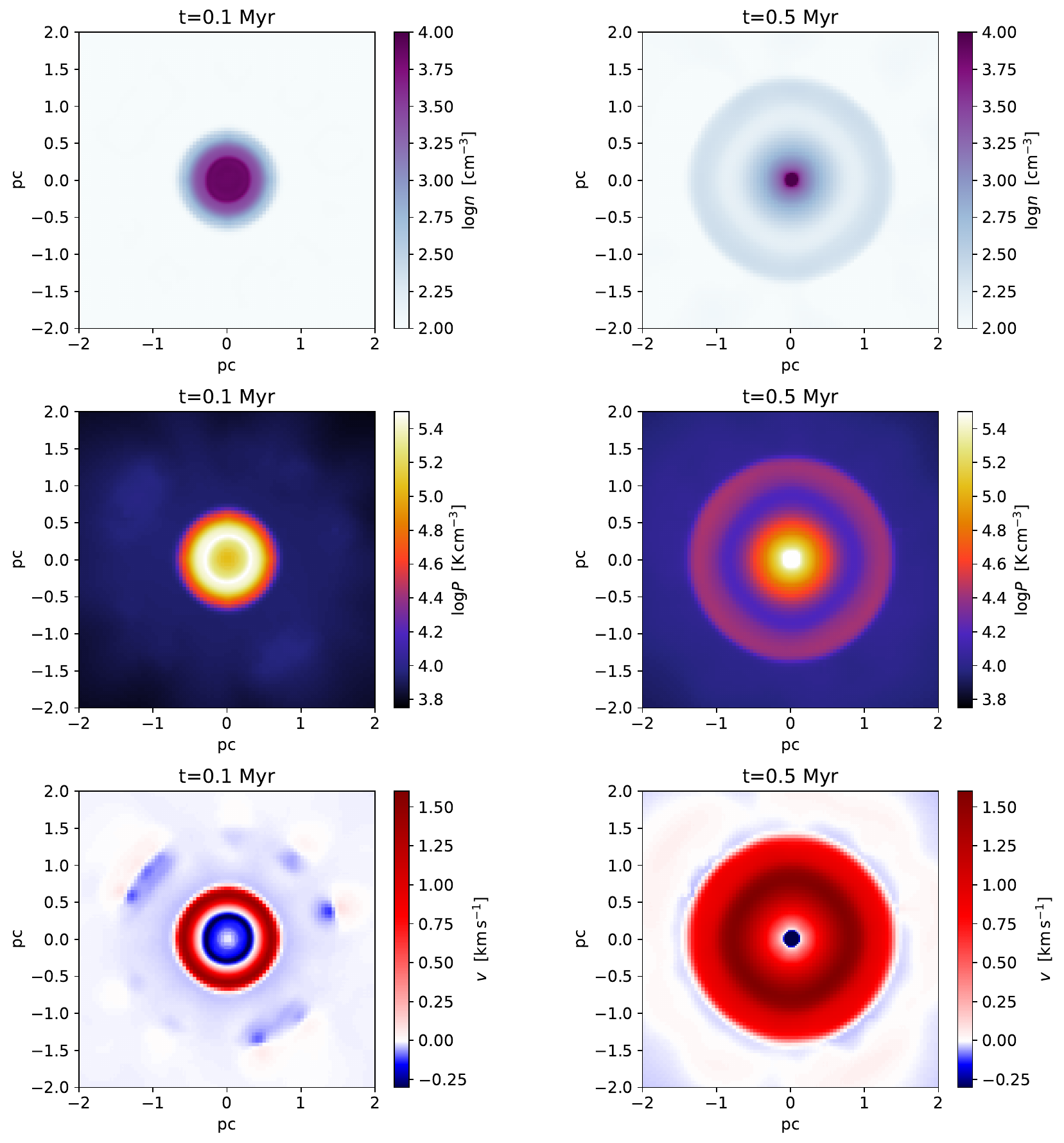}
\caption{Snapshots of the simulation \texttt{clump\_M50\_G2e3} at times $t=0.1$ Myr (left) and at $t=0.5$ Myr (right), obtained by slicing the computational box through its centre. {\bf Upper panels}: gas density (baryon number density $n$); for visualization purposes the colour range in the right-hand panel is reduced with respect to the maximum density, $n_{\rm max}\simeq 6\times 10^6\,\cc$. {\bf Central panels}: gas pressure normalized by the Boltzmann constant. {\bf Bottom panels}: gas radial velocity, with the convention that gas flowing towards the centre has negative velocity, and outflowing gas has positive velocity.
\label{clump_slice_M50_G2e3}
}
\end{figure*}

\begin{figure}
\centering
\includegraphics[width=0.49\textwidth]{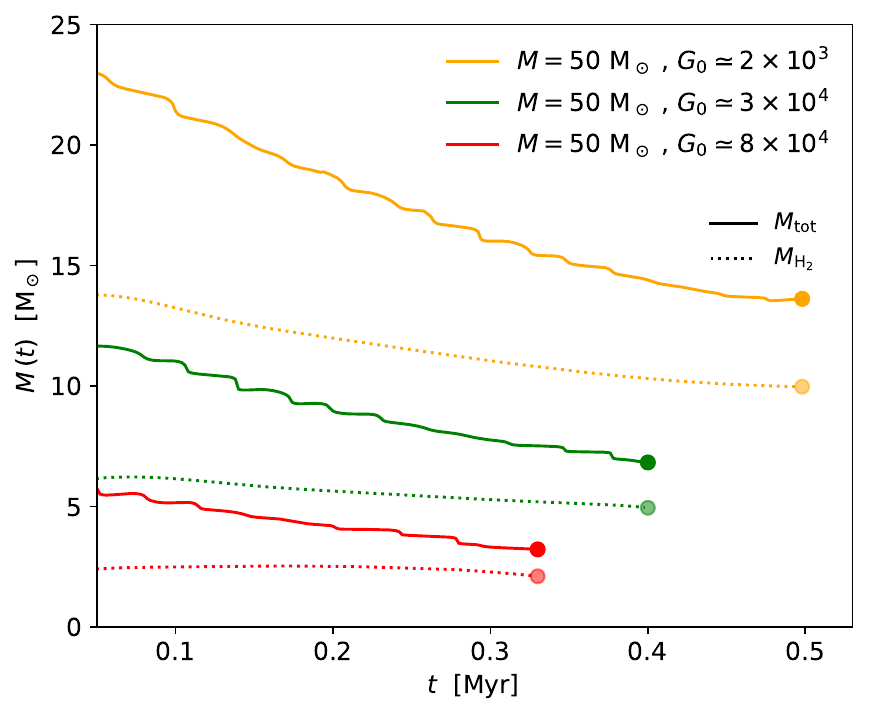}
\caption{Total clump mass (solid line) and molecular mass (dotted line), as a function of time, for the the runs of a 50 $\msun$ clump with radiation. The simulations are stopped when the clumps reach a minimum radius and gravitational collapse begins.
\label{clump_H2_M50}}
\end{figure}

We study first the evolution of a fiducial clump with initial mass of $50 \, \Msun$. The clump has an initial radius $R=0.5$ pc and a profile described by Eq. \ref{clump_profile} with a core density $n_c \simeq 6.6\times 10^3 \,\cc$. We run a simulation with no external radiation source, to be used as a control case.  We then perform three simulations introducing a nearly isotropic radiation field with different intensities $G_0$.

The clump radius is defined as the distance from the centre where the 99.7\% of the molecular gas mass is enclosed. In the run with no radiation, all the box is fully molecular, so we had to adopt a different definition of radius, namely where the $\h2$ density reaches 10\% of its maximum value at the centre.

\subsection{Run with no radiation}     %%%%

In Fig. \ref{clump_r_n_M50}, the clump without radiation sources is shown with grey lines: the plot tracks the evolution of the radius (left-hand panel) and the central density (right-hand panel).
The radius decreases by a factor of 100 during the collapse, leading to an increase of the maximum core density by 3 orders of magnitude. The clump reaches the minimum radius slightly before the free-fall time $t_\mathrm{ff}$ (shown by the vertical dotted line in the right-hand panel), and the maximum compression state is indicated by the grey circle. As opposed to works on protostellar accretion \citep{Larson1969, Shu1977}, we do not follow the clump evolution after the collapse, since our aim here is only to compare the implosion phase with that of clumps exposed to radiation.

\subsection{Runs with different impinging flux}     %%%%

We now want to compare the gravity-only simulation with the simulations where radiation is impinging on the clump surface. For the three simulations \texttt{clump\_M50\_G2e3}, \texttt{clump\_M50\_G3e4} and \texttt{clump\_M50\_G8e4}, the Habing flux $G_0$ through the clump surface is $\simeq 2 \times 10^3$, $3 \times 10^4$, and $8 \times 10^4$, respectively.

Slices of number density, thermal pressure and radial velocity for the $G_0=2\times 10^3$ run are shown in Fig. \ref{clump_slice_M50_G2e3}, at times $t=0.1$ Myr and $t=0.5$ Myr. In the first snapshot ($t=0.1$ Myr), it is possible to see that a thick layer of gas ($n\sim 500 \,\cc$) is pushed away at high velocity (Mach number $\mathcal{M}=1.5$), due to the underneath high-pressure gas ($P\sim 3\times 10^5 \,\mathrm{K}\,\cc$). This is due to radiation heating the clump surface, so that the pressure is increased and drives an expansion of an outer shell of gas. On the other hand, the overpressure (seen as the highest pressure layer in the first snapshot) also drives a shock inward. The shocked gas is flowing towards the centre at a velocity that is higher than the rest of the clump gas, which is undergoing gravitational collapse. The qualitative behaviour of the gas is the same found in \D17cit, albeit that work considered the effect of ionizing radiation but no gravity. Nevertheless, the effect of the FUV radiation is qualitatively the same: a surface layer of the clump is heated, so that it expands and drives a shock towards the centre of the clump. In the second snapshot, the clump is collapsed to a dense small core, while the photoevaporating flow has reached a higher distance from the centre.

\subsection{Radius and density evolution}\label{radius_density_evolution}    %%%% 

In Fig. \ref{clump_r_n_M50}, the evolution of clump radius and clump central density are shown for all the four runs. The radius decreases suddenly from the initial value, because of the photodissociation front moving into the cloud as an R-type front: the molecular hydrogen is thus photodissociated and the gas remains almost unperturbed, until the flux is attenuated (both because of photodissociation and dust absorption) and the front stalls. This phase lasts few kyrs (see Eq. \ref{DF_speed}). After that, the front drives a shock front compressing the gas ahead of it  \citep[D-type shock front,][]{Spitzer1998a}, so that the clump shrinks further because of the radiation-driven shock wave. Even if the flux is different in the 3 runs with radiation, the temperature at which the clump surface is heated does not vary much (200-250 K), hence the clump contraction proceeds almost at the same speed. 
The simulations are then stopped when the clump radius reaches a minimum of $\lsim 2 \Delta x_\mathrm{min}$, i.e. when the clump collapses to a size below the adopted resolution.

The right-hand panel of Fig. \ref{clump_r_n_M50} shows the clumps central density at different times. For the coloured lines, a circle marks the moment in which the clump has reached the minimum radius (i.e. about the size of a cell), and it corresponds to the higher compression state of the clump. The density reached in the implosion phase is lower when the radiation is stronger. This happens because for the simulations with strong flux the dissociated shell in the R-type phase is thicker, hence the remaining molecular collapsing core is less massive.

\subsection{Mass evolution}     %%%%

In order to compare the efficiency of the photoevaporation process with different radiation intensities, we plot in Fig. \ref{clump_H2_M50} the total clump mass $M_\mathrm{tot}$ (solid line) and the $\h2$ mass $M_{\mathrm{H}_2}$ (dotted line) as a function of time.

The lines start from the time $t_\textsc{r}$ which marks the end of the R-type dissociation front propagation. Subsequently, photodissociation  continues at the surface of the clump, generating a neutral flow such that both $M_\mathrm{tot}$ and $M_\h2$ decrease with time. Notice that the ratio $M_\mathrm{tot}/M_\h2$ is not constant, since the clump molecular fraction $x_{\h2}$ changes with time, even in the interior of the clump. 

In particular, $x_{\h2}$ is lower just after the R-type phase, when clumps have lower density. Indeed, both the bins [6.0, 11.2] eV (bin {\it a}) and [11.2, 13.6] eV (bin {\it b}, corresponding to Lyman-Werner band) play a role in the dissociation of $\h2$, as we have verified in some test simulations with only one radiation bin: the bin {\it b} dissociates the $\h2$ through the Solomon process, while the bin {\it a} causes an increase of the gas temperature, hence increasing the collisional dissociation of $\h2$. While the absorption of radiation in the bin {\it b} is very strong because of $\h2$ self-shielding, radiation in the bin {\it a} is basically absorbed by dust only\footnote{With the absorption cross sections adopted, the optical depth in the bin {\it a} is $\tau_a=1$ when $N_\textsc{h}=7\times 10^{20}\,{\rm cm}^{-2}$, while for LW radiation $\tau_b=1$ when $N_\textsc{h}=10^{15}\,{\rm cm}^{-2}$} and can penetrate even in the clump interior. At later times, clumps become denser and the gas cools more efficiently. As a result, $\h2$ formation is promoted, yielding the maximum value $x_{\h2}\simeq 0.76$ within the clump.

\subsection{Stability of the molecular core}     %%%%
\label{star_formation_in_the_clumps}

We have explicitly verified that the molecular core of the clumps is Jeans unstable when the simulations are stopped, implying that the clump could eventually collapse. This conclusion does not depend on the resolution of the simulation, as we have verified by running the simulation \texttt{clump\_M50\_G8e4} with $2\times$ and $4\times$ the standard resolution (see App. \ref{high_resolution}). Hence, the final clump mass is an upper limit to the final stellar mass $M_\star$. For the 50 $\msun$ clumps, we find $M_\star \simeq 15 \,\msun$ in the lowest flux case ($G_0=2\times 10^3$) and about $M_\star \simeq 3 \,\msun$ in the highest flux case ($G_0 = 8\times 10^4$). Following the standard Shu classification \citep{Shu1987, Andre1994, Andre2000}, star formation proceeds through an early accretion phase from the surrounding envelope ($t \simeq 1$ kyr), a late accretion phase via a disk ($t \simeq 0.2-1$ Myr) and a protostellar phase ($t \simeq 10$ Myr). We cannot investigate the effect of photoevaporation during these phases with the resolution and physics considered in our simulation suite, but we expect that the mass that goes into stars will be in general lower than $M_\star$ due to photoevaporating gas from the protostellar system. 

\section{Clumps with different masses}\label{Res_Mdiff}         % ------------------- SUBSEC : Clumps with different masses

\begin{figure*}
\centering
\includegraphics[width=\textwidth]{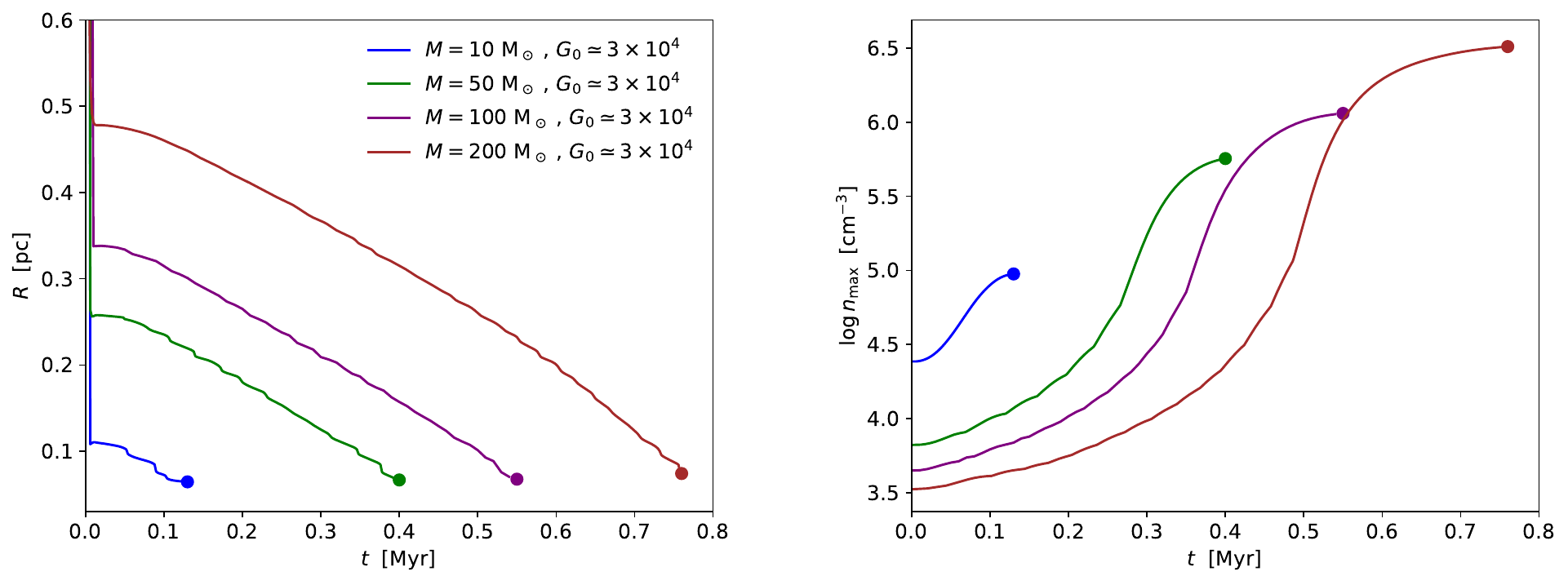}
\caption{Comparison of the evolution of the clump in the four simulations with the same impinging flux: \texttt{clump\_M10\_G3e4}, \texttt{clump\_M50\_G3e4}, \texttt{clump\_M100\_G3e4} and \texttt{clump\_M200\_G3e4}. {\bf Left}: variation with time of the clump radius, defined as in Fig. \ref{clump_r_n_M50}. {\bf Right}: variation with time of the clump maximum density, with the same symbols used in Fig. \ref{clump_r_n_M50}
\label{clump_r_n_Mdiff}}
\end{figure*}

\begin{figure*}
\centering
\includegraphics[width=0.47\textwidth]{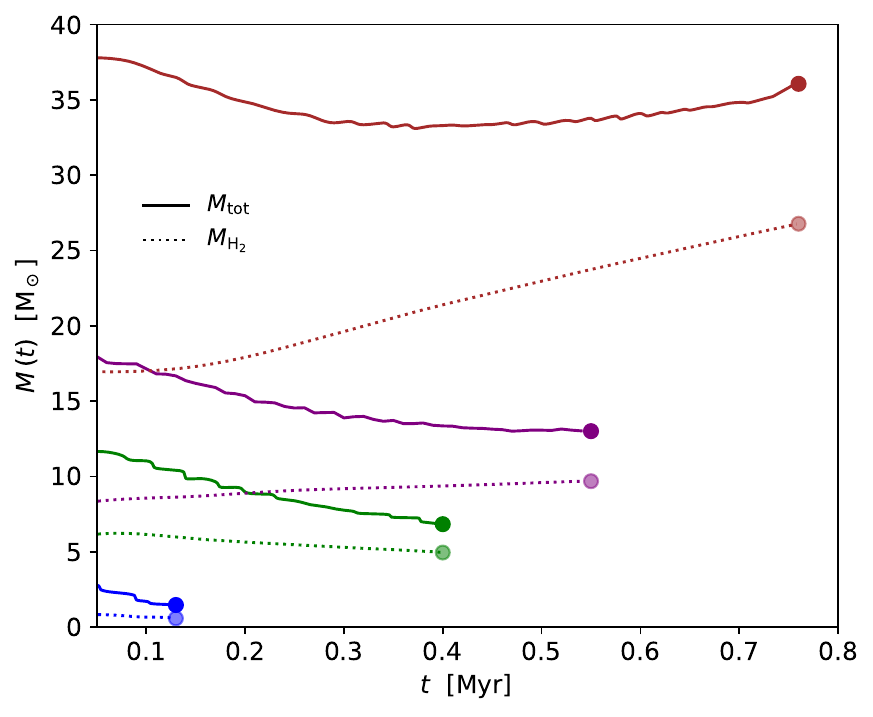}
\quad
\includegraphics[width=0.47\textwidth]{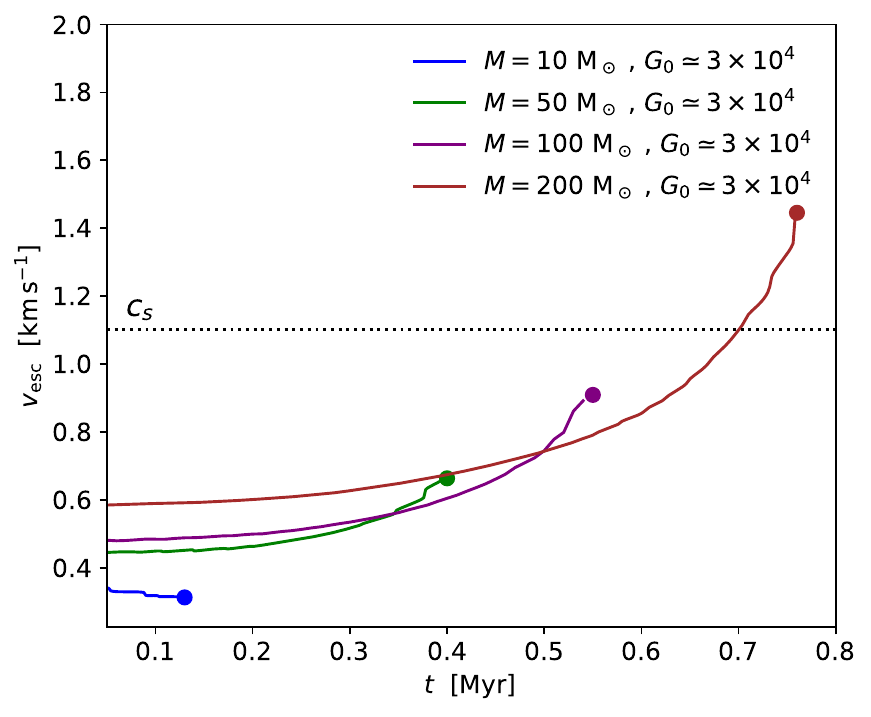}
\caption{
{\bf Left:} Total clump mass (solid line) and molecular mass (dotted line), as a function of time, for the the four simulations with the same impinging flux: \texttt{clump\_M10\_G3e4}, \texttt{clump\_M50\_G3e4}, \texttt{clump\_M100\_G3e4} and \texttt{clump\_M200\_G3e4}. The circles mark the time when the clump has reached the minimum radius in the simulations, and gravitational collapse begins.
{\bf Right:} Escape velocity from the surface of the molecular core, as a function of time. The dotted horizontal line marks the typical sound speed in the heat atomic layer of the clump, which approximates the speed of the photoevaporative flow.
\label{clump_H2_Mdiff}
}
\end{figure*}

After focusing on clumps with mass $50\,\msun$, we also analyse the effect of photoevaporation on clumps with lower (10 $\msun$) and higher (100 and 200 $\msun$) masses, with an initially impinging flux of $G_0\simeq 3 \times 10^4$ (see Tab. \ref{simulation_list}).

In the left-hand panel of Fig. \ref{clump_r_n_Mdiff} the clump radius is plotted as a function of time. Clumps with larger masses take more time to collapse, as they have a lower initial density and hence longer free fall time. The maximum density $n_\mathrm{max}$ as a function of time is shown in the right-hand panel, with higher values reached by more massive clumps.

In Fig. \ref{clump_H2_Mdiff} (left-hand panel), the total mass and the $\h2$ mass in the clumps are plotted. The $10\,\msun$ and $50\,\msun$ have the same trends seen in Sec. \ref{Res_M50}, with both total and $\h2$ mass decreasing during the implosion phase. Instead, the behaviour is different for the $100\,\msun$ and $200\,\msun$ clumps:
\begin{itemize}
\item $100\,\msun$: $M_{\h2}$ increases slightly (by 25\%)  during the implosion, while $M_\mathrm{tot}$ decreases slowly.
\item $200\,\msun$:  $M_{\h2}$ increases substantially (by 50\%) during the RDI, while $M_\mathrm{tot}$ does not decrease considerably and shows a slight raise after 0.4 Myr. 
\end{itemize}

The different behaviour of $M_{\h2}$ in the two more massive clumps, before the gravitational collapse, is due to the effect of radiation in the bin [6-11.2] eV. Photons in this band make their way to the clump interior, which is less dense in the centre with respect to smaller clumps, and increase the gas temperature so that the $\h2$ decreases in the centre. Nevertheless, when the clump collapses, the gas self-shields from this radiation and $\h2$ abundance increases again in the centre. This boost of the $\h2$ abundance compensates for the photoevaporative loss.

In the final part of the RDI, the two massive clumps behave differently, with $M_{\rm tot}$ raising for the 200 $\Msun$ clump. This is due to the fact that the escape velocity from the $200\,\msun$ becomes higher than the typical velocity of the outflowing gas. Fig. \ref{clump_H2_Mdiff} (right-hand panel) shows the time evolution of the escape velocity $v_\mathrm{esc}$ from the surface of the clump, for the four simulations with similar impinging flux. We have marked with a horizontal line the isothermal sound speed $c_s$ at $T\simeq 250$ K, which is a typical temperature of the heated atomic surface of the clumps. The circles mark the time when clump collapse. Only the 200 $\msun$ clump has $v_\mathrm{esc} > c_s$ at the end of the implosion phase, clarifying why this clump can accrete further in spite of the radiation impinging on its surface.

The same considerations of Sec. \ref{star_formation_in_the_clumps} hold: clumps are Jeans unstable at the end of the RDI, hence the remaining mass in the clump is an estimate of the mass going to form stars. Considering that photoevaporation can also remove mass from the protostellar system, such final mass should be regarded as an upper limit to the stellar mass. While the clump mass is generally reduced by photoevaporation during the RDI, due to its self-gravity the most massive clump in our set of simulations ($200\,\msun$) manages to retain its core mass after the R-type propagation of the DF. Hence in this case photoevaporation is completely inefficient in limiting star formation.

\section{General picture}\label{Res_final}         % ------------------- SEC : Final picture for photoevaporating clumps

\begin{figure*}
\centering
\includegraphics[width=\textwidth]{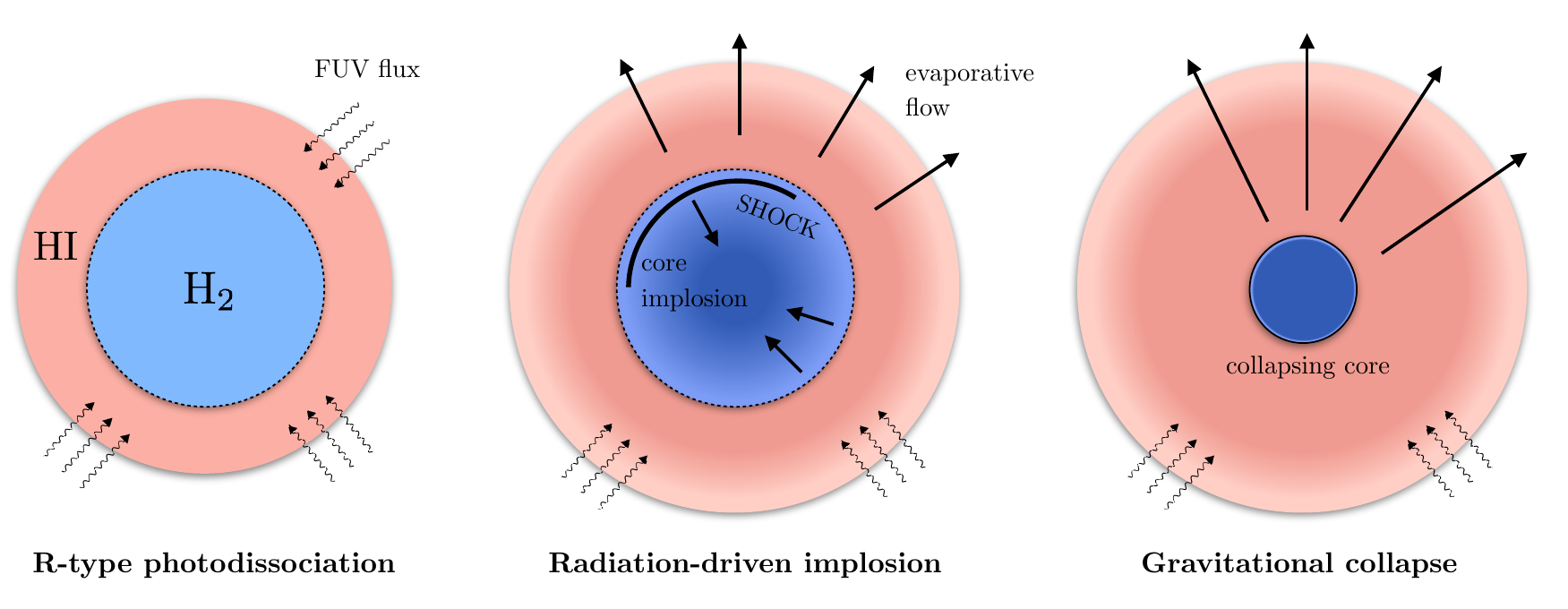}
\caption{Sketch of the 3 main phases of the photoevaporative process. From left to right: (1) FUV radiation penetrates as an R-type photodissociation front, turning a clump shell to atomic form, without any dynamical effect on the gas; (2) the clump undergoes an implosion phase, because of the high pressure of the photodissociated shell, while the atomic gas flows into the surrounding ISM; (3) the clump implodes to a Jeans unstable core, which undergoes gravitational collapse and star formation, if its mass is sufficient.  
\label{sketch_photoevaporation}}
\end{figure*}

\begin{figure*}
\centering
\includegraphics[width=1.02\textwidth]{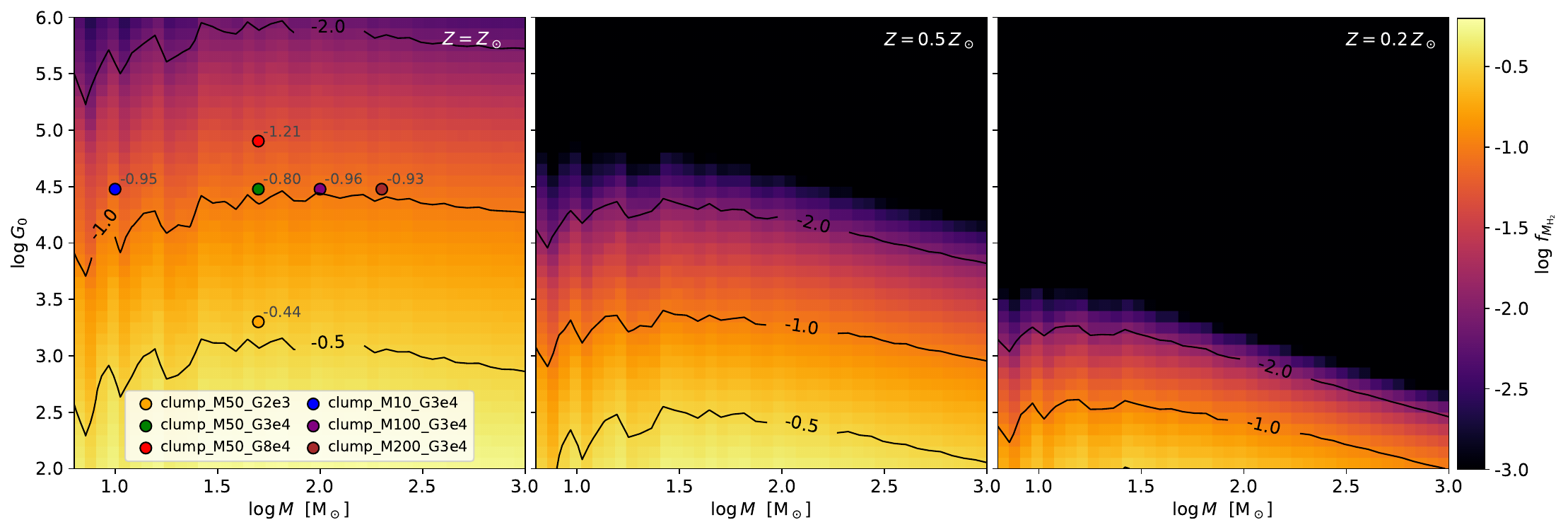}
\caption{Ratio of molecular mass after the R-type dissociation front propagation with respect to the initial value ($f_{M_\mathrm{H_2}}$), for different clump masses and different impinging FUV fluxes, obtained by running a set of 1D simulations. The three panels show the results for different gas metallicities ($Z=Z_\odot$, $0.5Z_\odot$, $0.2 Z_\odot$). In the first panel, the results from the 3D simulations are reported with dots, together with the corresponding $f_{M_\mathrm{H_2}}$ measured from the simulations.}
\label{clump_1d}
\end{figure*}

\begin{figure}
\centering
\includegraphics[width=0.49\textwidth]{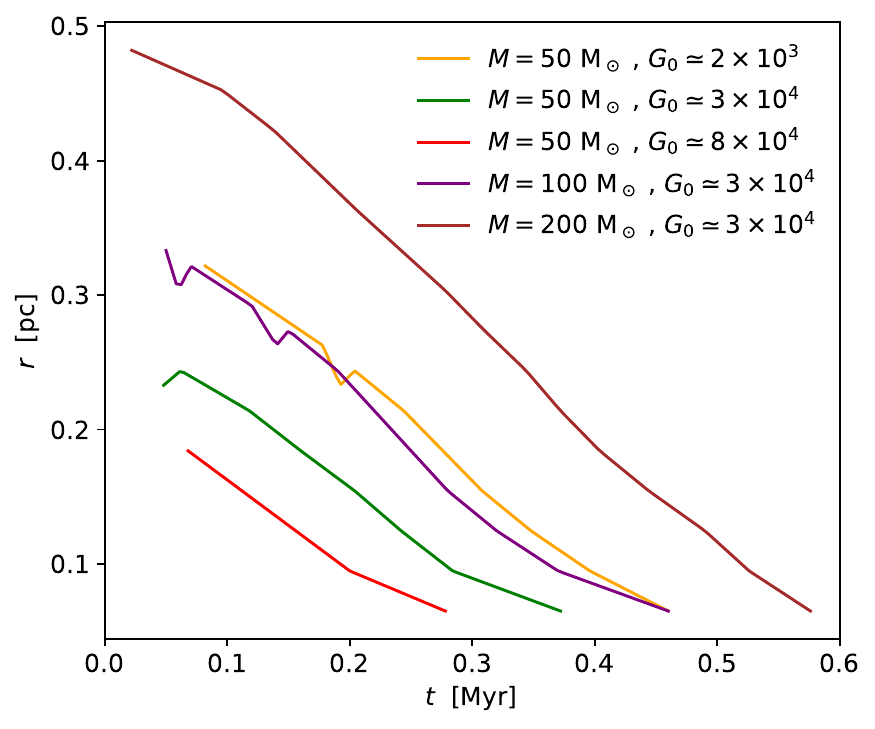}
\caption{The solid line tracks the shock position (radial distance from the centre of the clump) as a function of time, for the simulations \texttt{clump\_M50\_G2e3}, \texttt{clump\_M50\_G3e4}, \texttt{clump\_M50\_G8e4}, \texttt{clump\_M100\_G3e4}, \texttt{clump\_M200\_G3e4}. The shock position is found by producing the radial profile of velocity in the computational box and taking the position of the maximum negative velocity.
\label{clump_shock}
}
\end{figure}

The 3D simulations that we have run show that photoevaporating clumps undergo three main evolutionary phases, summarized in Fig. \ref{sketch_photoevaporation}:
\begin{itemize}
	\item[-] R-type DF propagation: FUV radiation penetrates the clump as an R-type front, with the clump density structure unaltered;
	\item[-] RDI: the heated atomic shell drives a shock inward, so that the clump implodes;
	\item[-] gravitational collapse: the molecular core is Jeans unstable, so it undergoes a gravitational collapse, with photoevaporation regulating the mass going into stars.
\end {itemize}
In the following, we analyse the details of these phases, trying to generalize the results of the simulations to a range of clump masses and impinging flux.

\subsection{R-type dissociation front propagation}

The first phase of the photoevaporative phenomenon is the dissociation of clump molecules by the propagation of the DF as an R-type front. During this phase, a clump shell is converted in atomic form during the DF propagation, without any dynamical effect on the gas. This phase is very short (less than 0.02 Myr), and it is responsible for the sudden decrease of $M_\h2$ with respect to its initial value, as we already pointed out in Fig. \ref{clump_H2_M50} and Fig. \ref{clump_H2_Mdiff}.

This phase determines substantially the fate of a clump. In fact, if the remaining $\h2$ mass is very small, the clump can be quickly eroded in the following photoevaporative phase. To make a prediction of the $\h2$ mass in the clump after the R-type DF propagation, we have run a set of 1600 1D simulations, with clump masses varying between 6 and $10^3 \, \msun$ and FUV fluxes in the range $G_0=10^{2-6}$. 
As initial setup of the 1D simulations, we have considered a 1D stencil passing through the centre of a three-dimensional clump. Hence, radiation is injected from both sides of this 1D box, with the prescribed flux.
The results are shown in Fig. \ref{clump_1d} (leftmost panel), where the colours mark the ratio of molecular mass after the R-type DF propagation to the initial molecular mass ($f_{M_{\h2}}$). With respect to the results of the 3D simulations, the 1D simulations slightly underestimate $f_{M_{\h2}}$, because of the different geometry. The plot shows that the fraction $f_{M_{\h2}}$ depends only weakly on the initial clump mass (apart from the low mass clumps, $M<30 \msun$), so that a relation between $f_{M_{\h2}}$ and $G_0$ can be derived by fitting the data:
\begin{equation}
\log(f_{M_{\h2}}) = -0.85 \log^2 G_0 + 0.22 \log G_0 - 0.38 
\end{equation}
We also notice that for the range of masses and fluxes that we have chosen, no clump is suddenly dissociated by the FUV radiation, hence all the analysed clumps will eventually undergo the implosion phase. 

We investigate the dependence of gas metallicity on the efficiency of photoevaporation. We run two additional sets of 1D simulations ($2 \times 1600$) with lower values of $Z$, i.e. $0.5 Z_\odot$, $0.2 Z_\odot$. The results for $f_{M_{\h2}}$ are shown in the central and right-hand panel of Fig. \ref{clump_1d}. Since the dust abundance is assumed to be proportional to $Z$, the low-$Z$ gas is more transparent to radiation, and molecular gas is dissociated more efficiently. Clumps are fully photodissociated by the R-type DF for $Z=0.5 Z_\odot$ ($Z=0.2 Z_\odot$) when $G_0>3\times 10^4$ ($G_0>3\times 10^3$). This finding agrees with results from \citet{Vallini2017} and \citet{Nakatani2018}, both pointing out that photoevaporation is more rapid in metal-poor clouds, i.e. the ones expect in high redshift galaxies \citep[e.g.][]{Pallottini2017a,Pallottini2019}.

\subsection{Radiation-driven implosion}

Clumps that are not completely dissociated by the propagation of the DF, will then attain a configuration with a molecular core surrounded by an atomic heated shell (this is the case for all clumps considered in this work). Since the latter has a higher pressure with respect to the molecular core, a shock propagates inward compressing the clump. This phase is generally called radiation-driven implosion.

In \D17cit we have developed an analytical solution for the propagation of the shock towards the centre of the clump. In that work, the shock parameters (as Mach number and compression factor) are computed as a result of the discontinuity between the cold molecular core and the heated atomic shell.
The backup pressure from the atomic shell is kept constant during the evolution, which is equivalent to assume a constant heating of the shell by radiation. Hence, the shock velocity $v_\mathrm{shock}$ changes only because of spherical convergence, causing an increase with radius as $v_\mathrm{shock} \sim r^{-0.394}$.
Nevertheless, this work shows that radiation is absorbed in the photoevaporative flow, hence the heating of the atomic gas is reduced and its pressure decreases accordingly. This effect acts in the opposite direction than the spherical convergence, reducing the shock speed.

In Fig. \ref{clump_shock}, the shock radial distance from the centre is plotted as a function of time. The simulation \texttt{clump\_M10\_G3e4} has been excluded because the resolution is too low to resolve the shock position properly. The initial position of the shock is determined by the clump radius at the end of the DF propagation (cfr. left-hand panel of Figs \ref{clump_r_n_M50} and \ref{clump_r_n_Mdiff}). The plot shows that the shock speed is almost constant in time, implying that there is a balance between absorption of radiation which is backing up the shock, and spherical convergence of the shock. The simulations show that the shock Mach number (with respect to the gas ahead of the shock) is $\mathcal{M}=2.0 \pm 0.3$, with a shock speed about half of the speed of sound in the heated clump shell ($c_\textsc{pdr}$). This is in contrast to \citet{Gorti2002} assumption, where the shock speed is approximated to be exactly $c_\textsc{pdr}$.

Finally, we notice that the RDI lasts more than the time needed for the radiation-driven shock to reach the centre of the clump, in contrast with \citet{Gorti2002} assumption that the clump stops contracting when the shock has reached the centre. This is evident especially for massive and larger clumps, and it is due to the fact that the shock moves towards the centre with a higher speed than the collapsing clump surface. Thus, when the shock has reached the centre, the surface is still moving inward.

During the RDI, two effects changing the $\h2$ mass are competing: (1) the clumps loses mass from its surface, (2) the central density increases, raising the $\h2$ abundance. The second effect dominates for clumps massive enough ($M \geq 100 \,\msun$). Furthermore, self-gravity inhibits the photoevaporative flow of even more massive clumps ($M\geq 200\,\msun$), showing that in this case photoevaporation is not effective in reducing the total clump mass during the RDI.

\subsection{Gravitational collapse}

After the RDI, the molecular core is still Jeans unstable, thus we expect it to undergo a gravitational collapse with possible star formation (if the molecular mass of the core is sufficient). We do not include star formation routines in our simulations, as done for instance via seeding of a protostellar object and by following its accretion \citep{Dale2007, Peters2010}.

Other works on photoevaporation \citep{Gorti2002, Decataldo2017} do not include clump self-gravity in their analysis.  As a result, in those works the clump reaches a minimum radius where thermal pressure balances the pressure of the shell heated by radiation. In this scenario, gas continues to photoevaporate from its surface, until all the molecular gas is dissociated. This is not seen in our simulations, since gravity leads to the clump self-collapse after the RDI.

The clump mass at the end of the R-type DF propagation (Fig. \ref{clump_1d}) is in general an upper limit to the mass  $M_\star$ going into stars. In fact during the RDI, a fraction of the mass flows away from the clump. However, this does not happen for clumps with sufficiently large masses in which self-gravity prevents the gas from escaping the clump. 

% ************************************************************************************************ CONCLUSIONS

\section{Conclusions}\label{Con}

We have studied the photoevaporation of Jeans unstable clumps by Far-Ultraviolet (FUV) radiation, by running 3D RT simulations including a full chemical network which tracks the formation and dissociation of $\h2$.

The simulations have been run with the adaptive mesh refinement code $\ramses$ \citep{Teyssier2002} by using the $\ramsesrt$ module \citep{Rosdahl2013}, in order to perform momentum-based on-the-fly RT. The RT module has been coupled with the non-equilibrium chemical network generated with $\krome$ \citep{Grassi2014}, in order to consider photo-chemical reactions, as the dissociation of $\h2$ via the two-step Solomon process.
We have run seven simulations of dense clumps, embedded in a low density medium ($n=100\,\cc$), with different clump masses ($M=10-200\,\msun$) and different impinging FUV radiation fields ($G_0=2\times 10^3-8\times 10^4$). These clumps have central number densities $n_c\simeq 6\times 10^3 - 2\times 10^4$ and total column density $N_{\textsc{h}_2}\simeq 5\times 10^{21} {\rm cm}^{-2}$.

In all the cases, we find that the evolution a clump follows three phases:
\begin{enumerate}
\item[{\bf 1)}] R-type DF propagation: the density profile remains unaltered while most of the $\h2$ mass is dissociated ($40-90\%$ of the $\h2$ mass, depending on $G_0$).

\item[{\bf 2)}] RDI: the heated shell drives a shock inward ($\mathcal{M}\simeq2$) promoting the clump implosion; at the same time, the heated gas at the surface evaporates with typical speed $1.5-2\,\kms$.

\item[{\bf3)}] Gravitational collapse of the core: the clump collapses if the remaining $\h2$ core is Jeans unstable after the RDI; if $M_\h2$ is significantly higher than 1 $\msun$, than we expect it to form stars.
\end{enumerate}

During the RDI, both the molecular mass $M_{\textsc{h}_2}$ and the total mass $M$ decrease for the 10 and $50\,\msun$ clumps. However, we find that $M_{\textsc{h}_2}$ tends to increase by $\sim 25-50\%$ for more massive clumps, due to the fact that previously dissociated $\h2$ recombines when the clump collapses and the density increases. For the most massive clump only ($200\,\msun$), photoevaporation is inefficient even in reducing the total mass $M$, since during the RDI the escape velocity becomes larger than the outflowing gas speed (comparable to the $\hi$ sound speed).

All the $\h2$ cores are still Jeans unstable after the RDI. This shows that FUV radiative feedback is not able to prevent the gravitational collapse, although it regulates the remaining molecular gas mass. All the simulated clumps manage to retain a mass $M>2.5\,\msun$, hence suggesting that star formation may indeed take place. The evolution of low-mass clumps follows what expected from analytical works \citep{Bertoldi1989, Gorti2002, Decataldo2017}. However, our analysis clarifies that self-gravity has a non negligible effect for massive clumps ($\apprge 100 \msun$), limiting the mass loss by photoevaporation.

The dynamics of photoevaporating clumps can also have important consequences for their Far-Infrared (FIR) line emission \citep{Vallini2017}, \CII~in particular. In fact, a strong $G_0$ increases the maximum \CII~luminosity, as FUV radiation ionizes carbon in PDRs, albeit short-lived clumps may contribute less significantly to the parent GMC emission. To understand the effect of photoevaporation on line luminosity, simulations accounting for the internal structure of GMCs are required. This would allow to track the contribution of many clumps with different masses and subject to different radiation fields. We will address this study in a forthcoming paper.

Finally, we point out that photoevaporation is also a crucial effect in regulating the molecular mass in ultra-fast outflows launched from active galaxies. Indeed, dense molecular gas ($n\sim 10^{4-6} \, \cc$) is observed up to kpc scale \citep{Cicone2014, Bischetti2018, Fluetsch2019}, and its origin and fate are currently under investigation \citep{Ferrara2016c, Decataldo2017, Scannapieco2017, Richings2018}. 
In addition to FUV radiation, these clumps are also subject to a strong EUV field (not considered in this work), causing a fast photoevaporative flow of ionized gas and a stronger RDI. Since radiation is coming from the nuclear region of the active galaxy, the radiation field seen by the clump is non-isotropic, affecting only the side of the clump facing the source. Moreover, clump masses are generally larger ($10^{3-4}\,\msun $, see \citealt{Zubovas2014a, Decataldo2017}) and high turbulence is reasonably expected within the outflow.
Nevertheless, clumps will still follow a similar evolutionary path, with R-type dissociation/ionization, RDI, and gravitational collapse if the imploded core is Jeans unstable. This can be relevant to explain the observed molecular outflow sizes, which depends on the lifetime of clumps undergoing photoevaporation. 
Moreover, in this scenario the recent hints of star formation inside the outflow \citep{Maiolino2017, Gallagher2019, RodriguezdelPino2019} seem plausible, in view of our finding that most clumps manage to retain a sufficient amount of mass to collapse and form stars.

% ************************************************************************************************ ACKNOWLEDGMENTS

\section*{Acknowledgments}

This research was supported by the Munich Institute for Astro- and Particle Physics (MIAPP) of the Deutsche Forschungsgemeinschaft (DFG) cluster of excellence \quotes{Origin and Structure of the Universe}.
We thank A. Lupi, J. Rosdahl, and the participants of the \quotes{The Interstellar Medium of High Redshift Galaxies} MIAPP conference for fruitful discussions.
The simulations have been run on the UniCredit Research \& Development facilities; in particular we thank M. Paris for his technical support.
DD and AF acknowledge support from the European Research Council (ERC) Advanced Grant INTERSTELLAR H2020/740120.
LV acknowledges funding from the European Union's Horizon 2020 research and innovation program under the Marie Sk\l{}odowska-Curie Grant agreement No. 746119. 
We acknowledge use of the $\textsc{python}$ programming language,  $\textsc{astropy}$~\citep{AstropyCollaboration2013},  $\textsc{matplotlib}$~\citep{Hunter2007},  $\textsc{numpy}$~\citep{VanderWalt2011}, \pymses~\citep{Labadens2012}.

% ************************************************************************************************ BIBLIOGRAPHY

\bibliographystyle{style/mnras}
\bibliography{library}

% ************************************************************************************************ APPENDIX

\FloatBarrier % enabled by package placeins
\appendix

\section{Benchmarks}\label{app_tests}

\begin{figure}
\centering
\includegraphics[width=0.49\textwidth]{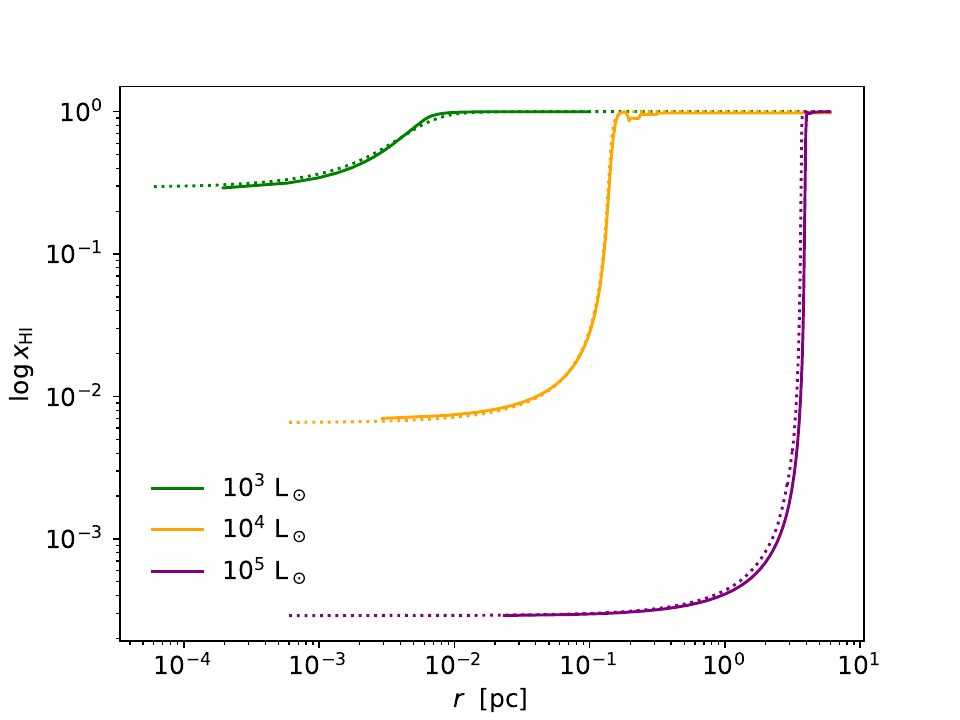}
\caption{Neutral fraction as a function of depth, for a 1D gas slab with radiation coming from the left side. The source is a black body at a distance of 1 pc, and different colours stand for different bolometric luminosities. The solid lines show the results obtained with a simulation using the code $\ramsesrt$ coupled with $\krome$, while the dotted lines are obtained with a semi-analytical model. See App. \ref{app_hii} for details.
\label{test_hii}}
\end{figure}

\begin{figure}
\centering
\includegraphics[width=0.49\textwidth]{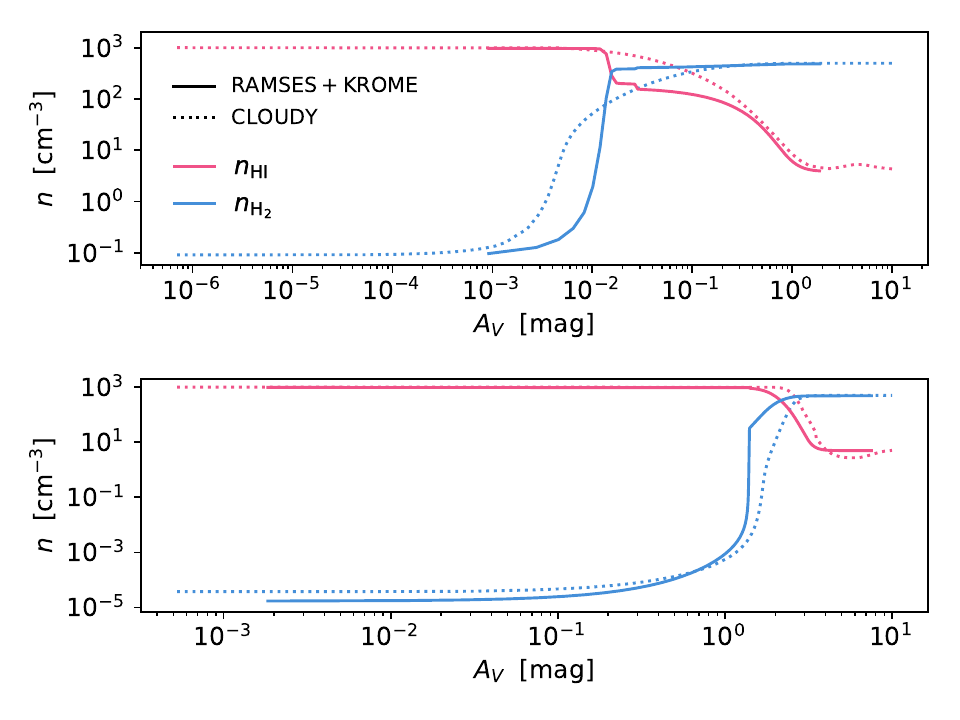}
\caption{Atomic (red) and molecular (blue) hydrogen density as a function of visual extinction, for a PDR-like gas slab with FUV radiation coming from the left side. The solid lines show the result obtained with $\ramsesrt$ coupled with $\krome$, while the dotted lines show the result obtained with $\textsc{cloudy}$ in the context of PDR code comparison \citep{Rollig2007}.
\label{test_pdr}}
\end{figure}

\begin{figure*}
\centering
\includegraphics[width=\textwidth]{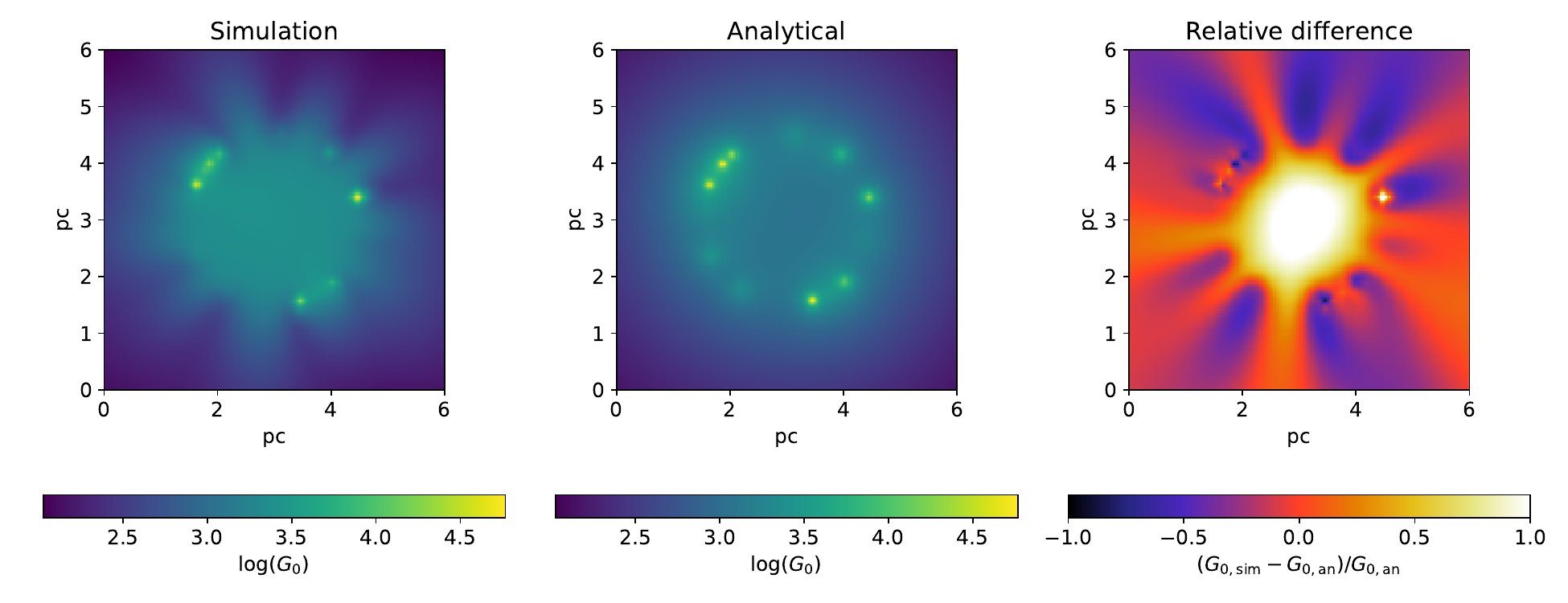}
\caption{Comparison between the radiation field obtained in the simulation and the radiation field obtained analytically, for a domain with homogeneous gas density (100 $\cc$) and no dust. {\bf Left}: Total flux in the FUV band in a slice of the computational domain. {\bf Middle}: analytical result for the same configuration of the simulation. {\bf Right}: relative difference between simulated and analytical radiation fields.
\label{radiation_test}}
\end{figure*}

\begin{figure}
\centering
\includegraphics[width=0.49\textwidth]{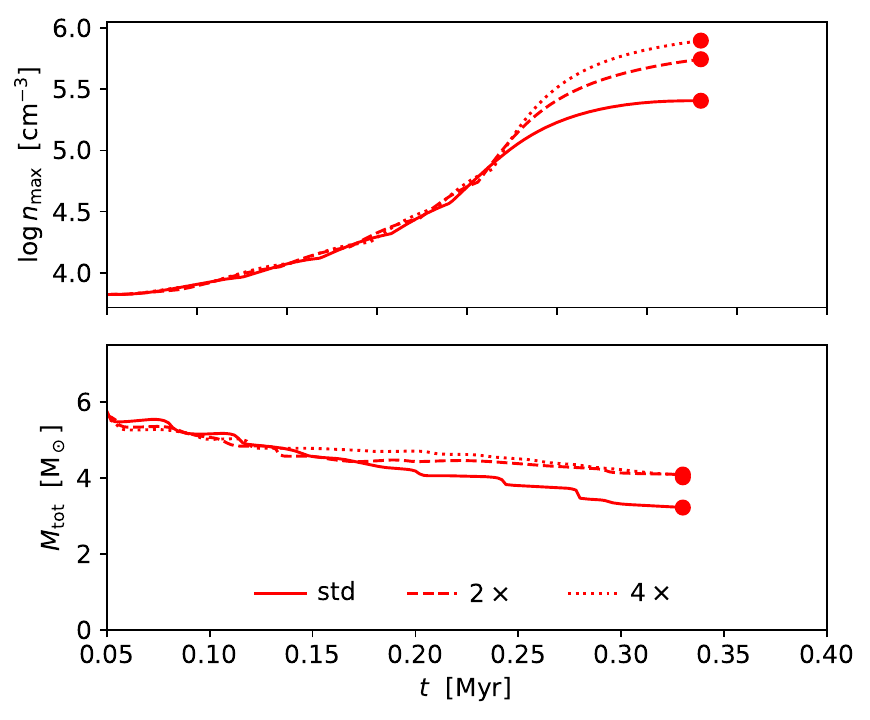}
\caption{Maximum density in the clump (upper panel) and total clump mass (lower panel) as a function of radius, for a $50\,\msun$ clump subject to a FUV radiation field with $G_0=8\times 10^4$. The three lines are obtained by running the simulation with different maximum resolution, as detailed in App. \ref{high_resolution}.
\label{clump_res}}
\end{figure}

\subsection{\HII region}\label{app_hii}

As a first test of our version of $\ramsesrt$ coupled with $\krome$, we run a simulation of 1D slab of atomic gas invested by ionizing radiation. The test is aimed to check the coupling between radiative transfer and chemistry, hence we run in static mode, i.e. forcing the gas velocity in every cell to zero. The 1D computational domain has a size $L=0.2$ pc, resolved with 1024 cells, and it is filled with neutral hydrogen, at a density of 100 $\cc$.

The radiation field comes from the left side of the slab, with the spectrum of a black body source at a distance of 1 pc, discretised in 10 bins (delimited by 13.6, 18.76, 23.9, 29.1, 34.2, 39.4, 44.5, 49.8, 54.8, 60.0, 1000.0 eV). The geometrical flux reduction is not considered, while dust cross section is set to $\sigma_d = 2.94\times 10^{-22} \, \mathrm{cm}^2$. We note that dust is not relevant since the column density in the whole slab is lower than $10^{20}\, \mathrm{cm}^{-2}$. The simulations are stopped at 1 kyr, i.e. longer than the recombination time $t_\mathrm{rec}\simeq n/ \alpha_\textsc{b}$.

The result is compared with a semi-analytical solution, obtained by dividing the slab of gas in 1024 cells and proceeding from the first cell to the $n$-th cell as follows:
\begin{enumerate}
\item[1.] compute the temperature $T$ in the $n$-th cell, using heating and cooling function provided by \citet{Gnedin2012}; 
\item[2.] compute the ionization fraction $x_\textsc{}(T)$ in the $n$-th cell, according to 
\be
x(t)= \left(   \dfrac{1}{n \alpha_\textsc{b}(T)}  \int_{13.6\, eV}^\infty F_\nu \, a_\nu \, \mathrm{d}\nu \right)^{1/2}
\ee
where $n$ is the H nuclei number density, $\alpha_\textsc{b}(T)$ is the case B recombination coefficient \citep[taken from][]{Abel1997}, $F_\nu$ is the specific flux at the cell and $a_\nu$ is the photoionization cross section \citep[values in][]{Verner1996}.
\item[3.] compute the total \HI column density $N_\textsc{hi}$ from cell $1$ to $n$
\item[4.] reduce radiation according to absorption due to $N_\textsc{hi}$
\end{enumerate}

The comparison between the simulation and the semi-analytical model is shown in Fig. \ref{test_hii}, showing the neutral fraction $x_\textsc{hi}$ as a function of the depth into the gas slab, for black bodies sources with different bolometric luminosities ($10^3$, $10^4$, $10^5$ $\lsun$).
The solid lines shows the results of the simulations, while the dotted lines are obtained with the semi-analytical model. The curves for the $10^3$ and $10^4 \, \lsun$ are in very good agreement, while there is a small difference for the the $10^5 \,\lsun$ source near the Str\"omgren depth $\delta_{\rm Str}$.
We have verified that this is due to slightly different gas temperatures around $\delta_{\rm Str}$, which in turn implies a difference in the recombination coefficient. Indeed, $\krome$ and \citet{Gnedin2012} do not use the same coefficients for computing heating and cooling functions.

\subsection{PDR}\label{app_pdr}

In order to test the chemistry of molecular hydrogen, we have carried out a 1D simulation of a slab of molecular gas, with radiation coming from one side. The setup is the same of typical simulations carried out with 1D time-independent PDR codes, as $\textsc{kosma}$-$\tau$ \citep{Stoerzer1996}, $\textsc{cloudy}$ \citep{Ferland1998}, $\textsc{ucl\_pdr}$ \citep{Bell2005} and many others. A PDR-code comparison has been done by \citet{Rollig2007}, that benchmarks the results obtained with the different codes starting from the same setup conditions.

In particular, here we compare the results of our code with the models V1 and V2 in the R\"ollig work. In the V1 and V2 models, the gas filling the computational domain has an H density $n_\mathrm{H}=10^3\,\cc$, with He elemental abundance $A_\mathrm{He}=0.1$. The flux is coming from the left side, and it is given by $\chi$ times the Draine spectrum \citep{Draine1978} in the FUV (6.0 - 13.6 eV); the V1 model has $\chi=10$, while V2 has $\chi=10^5$.

The dust cross section is $\sigma_d = 1.75 \times 10^{-21} \,\mathrm{cm}^2$ and does not depend on frequency, the cosmic ray H ionization rate is $\zeta_\mathrm{H}=5\times 10^{-17}\, \mathrm{s}^{-1}$ and the $\h2$ formation and dissociation rates are set to $R_\h2 =3 \times 10^{-18} T^{1/2}$ and $\Gamma_\h2 = 5.10\times 10^{-11}\, \chi)$ respectively. Note that in \citet{Rollig2007}, such prescription on the dissociation rate is adopted only for the codes that do not compute explicitly it by summing over all oscillator strengths.

We prepare our simulation to match the setup of the V1 and V2 models in the PDR benchmark, in particular by setting the same $\h2$ formation and dissociation rate of molecular hydrogen. The main drawbacks with respect to PDR codes are that (1) we use only the two FUV bins [6.0-11.2] eV and [11.2-13.6] eV to sample the Draine spectrum, and (2) we do not track C, O and Si as separate species.

The results of the comparison are displayed in Fig. \ref{test_pdr}, for model V1 (upper panel) and model V2 (lower panel). The $\h2$ number density and the $\hi$ number density are represented with a blue and a red line respectively; with solid lines we plot the result from our $\ramsesrt$ + $\krome$ simulation;  with dashed lines we plot results\footnote{Data from \citet{Rollig2007} is available at \url{http://www.ph1.uni-koeln.de/pdr-comparison}.} for the $\textsc{cloudy}$ test. 
The depth into the gas slab in the x axis is measured as a visual extinction $A_V$, which is proportional to the H column density, according to $A_V = 6.289\times 10^{-22} \, N_H$. We obtain a good agreement with $\textsc{cloudy}$, matching the $\hi$ and $\h2$ abundances at low/high $A_V$. The small difference in the residual $\h2$ abundance in the V2 test is due to a difference in the temperature, affecting the $\h2$ formation rate. The position of the $\hi-\h2$ transition is well captured in the simulations, with a steeper transition profile due to the different $\h2$ self-shielding recipe.

\section{Opposite colliding beams problem}\label{app_oppositeflux}

Radiative transfer codes based on the M1 closure relations give unexpected results when radiation beams that travel in opposite directions collide \citep{Rosdahl2013}. The flux in the computational domain can be different to what expected by summing the flux of the single beams, showing an excess in the direction perpendicular to the beams.

With the purpose of testing the behaviour of the radiation field for the setup of our simulations, we have run a simulation with resolution $128^3$, with 50 stars with luminosity $10^4 \, \lsun$ at a distance of 1.5 pc from the box centre and constant gas density inside the whole domain ($n=10^{-3}\,\cc$). Dust is not included and gas absorption is negligible at this density, so the Habing flux can be easily calculated analytically for comparison: the flux in every cell is computed by summing the contribution to the flux from all sources, by scaling it with the square root of the distance.

The result after a time sufficient for radiation to cross the whole domain is showed Fig. \ref{radiation_test}, where flux in the FUV band (left-hand panel) is compared with the analytical solution for the same setup (middle panel). From the map of the relative difference between the two (right-hand panel), we observe that there is a factor of 2 at most. Nevertheless, in the region inside the spherical shell where stars reside, the flux is rather homogeneous. Hence, despite the opposite colliding beams problem, we can carry out our simulations accounting for the fact that the flux on the clump is higher than what expected from an analytical prescription.

\section{Convergence test}\label{high_resolution}

The standard resolution of our simulation suite is $\Delta x \simeq 0.047$ pc, refining up to $\Delta x \simeq 0.023$ pc according to a $\h2$ density gradient criterion. Due to radiation-driven implosion and gravitational instability, all the clump mass collapses to one cell at some time $t_c$ which depends on the initial mass and the intensity of the radiation field. We stop our simulations at $t_c$, since the later time evolution of the clumps would be unresolved.

We argue that -- despite the resolution limit -- $t_c$ is a good estimate of radiation-driven implosion duration and, being the final clump Jeans unstable. Its mass at $t_c$ represents an upper limit to the gas mass that collapses and forms stars.
To show this point, we perform a convergence test: we take the $50\,\msun$ (with $G_0=8\times 10^4$) clump as a reference and we run the two additional simulations: (1) we increase the resolution by a factor 2 in a central region with radius 0.1 pc; (2) we add a further level of refinement in a region with radius 0.05 pc. 

Fig. \ref{clump_res} shows the time evolution of the maximum density (upper panel) and the clump mass (lower panel) for the runs with different resolution. Simulations are stopped at the time $t_c$ when the clump reaches the minimum resolvable size (1-2 cells): $t_c$ is roughly the same in the three cases.
In fact, by increasing the resolution, we can resolve the clump collapse to a later stage, but since the timescale for collapse becomes shorter, the clumps reach the minimum size in almost the same time. The lower panel confirms that the final clump mass is not affected by numerical resolution, implying that the estimate of the clump mass at the end of the RDI is reliable.

\label{lastpage}
\end{document}